\documentclass[11pt,a4paper]{article}

\usepackage{fontspec}
\usepackage[a4paper, margin=25mm]{geometry}
\usepackage{setspace}
\usepackage{graphicx}
\usepackage[colorlinks=true,linkcolor=blue,citecolor=blue,urlcolor=blue]{hyperref}
\usepackage{amsmath,amssymb}
\usepackage{booktabs}
\usepackage{caption}
\usepackage{float}
\usepackage{parskip}
\usepackage{titlesec}
\usepackage{fancyhdr}
\usepackage{microtype}

\titleformat{\section}{\large\bfseries}{\thesection.}{0.5em}{}
\titleformat{\subsection}{\normalsize\bfseries}{\thesubsection}{0.5em}{}

\graphicspath{{./figures/}}

\begin{document}

\thispagestyle{fancy}

\begin{center}
{\LARGE\bfseries The Digital Afterlife of Empires:\\[4pt]
Four Language Models Converge on\\[4pt]
the Same Imperial Cartography of Writing}

\vspace{20pt}

{\large Hiroki Fukui, M.D., Ph.D.}

\vspace{8pt}

{\small
ORCID: 0009-0008-7122-522X\\[4pt]
Research Institute of Criminal Psychiatry / Sex Offender Medical Center\\
Department of Neuropsychiatry, Kyoto University\\[4pt]
Email: fukui@somec.org\\[8pt]
\textit{Part II of the Kotonoha Series. Companion paper: Fukui (2026), arXiv:2604.10957 (q-bio.PE).}
}

\vspace{30pt}
\end{center}

\section*{Abstract}

Large language models process the world's writing systems with radical
inequality. We constructed the Digital Script Representation Index
(DSRI), a seven-axis measure spanning tokenizer efficiency, Unicode
standardization, web corpus volume, LLM knowledge accuracy, optical
character recognition, machine translation, and input method
availability. Applied to the 300 writing systems of the Global Script
Database (Fukui, 2026), the DSRI reveals that only 29 scripts (9.7\%)
are fully supported by contemporary digital infrastructure. Among 158
living scripts, 60 (38.0\%) lack complete support. Tokenizer efficiency
varies by a factor of 31.7 across 45 scripts measured with parallel
text. A serial mediation model --- imperial intervention → speaker
population → web corpus → tokenizer efficiency --- is consistent with
full mediation, with the direct effect of empire indistinguishable from
zero (β = −0.22, p = 0.39), an incremental ΔR$^{2}$ of 0.366 contributed by
digital presence, and structural equation model fit indices
indistinguishable from a saturated model at n = 45 (CFI = 1.01, RMSEA =
0.00); the bias-corrected bootstrap CI on the indirect effect grazes
zero, and we treat the mediation evidence as suggestive rather than
confirmatory (§3.3, §4.5). Across four independent LLM families (Claude,
GPT-4o, Grok, DeepSeek; 12,000 API calls), base-rate-deviation error
patterns converge at Spearman ρ = 0.85--0.98 (all p \textless{} 0.002).
One hundred seventy-two script--feature items are answered identically
wrong by all four models; over-attribution outnumbers under-recognition
3.9:1, and the single feature ``used for religion'' concentrates 43.6\%
of convergent errors (enrichment ratio 4.1×). When religion is excluded
as a sensitivity check, the cross-architecture convergence is preserved
(mean ρ = 0.87 on the remaining nine features) and the over-attribution
asymmetry persists at 1.77:1 (n = 97, binomial p = 0.008), indicating
that the convergent bias is multi-channeled rather than
single-channeled. The findings are consistent with an interpretation in
which the structural inequalities historical empires inflicted on script
communities persist in contemporary language models through the shared
training corpus rather than through any individual model's design
choices.

\section{Introduction}

In 2016, the Unicode Consortium encoded Adlam, a script invented in the
1980s by two Guinean teenagers to write the Fulani language. An
estimated 40 million people across West Africa use Adlam today. A decade
after its encoding, every major large language model tokenizer processes
Adlam characters by decomposing them into raw UTF-8 byte sequences ---
the computational equivalent of treating each letter as meaningless
noise. An Adlam user querying an LLM pays, in tokenized units, roughly
twenty times what an English user pays for the same semantic content.
This disparity was not designed. No engineer decided to disadvantage
Fulani speakers. And yet the structure that produces it is not new; it
is five centuries old.

In a companion paper (Fukui, 2026), we demonstrated that writing systems
evolve with clock-like regularity --- accumulating structural changes at
a measurable rate of q = 0.226 substitutions per character per
millennium --- and that political interventions, particularly imperial
conquests, break this clock. The Global Script Database (GSD) documented
300 writing systems across 43,000 years and showed that the Spanish
Empire destroyed 50\% of the scripts it contacted, that colonial-era
extinction rates were 4.2 times background, and that the magnitude of
structural deviation from the molecular clock quantifies the violence
imposed on a script community. That paper asked what empires did to
writing. This paper asks whether they have stopped.

The question is motivated by an observation that is simple to state and
difficult to dismiss: the scripts that large language models process
most efficiently are the scripts of historical empires, and the scripts
they process least efficiently are the scripts those empires suppressed.
This correlation could be trivial --- a mere reflection of present-day
speaker populations and internet penetration --- or it could be
structural, tracing a causal chain from imperial violence through
demographic collapse, digital corpus scarcity, and tokenizer design back
to the economic exclusion of the communities that survived.
Distinguishing between these possibilities requires measurement.

We constructed the Digital Script Representation Index (DSRI), a
seven-axis instrument that quantifies the digital presence of each of
the 300 scripts in the GSD across tokenizer efficiency, Unicode
standardization, web corpus volume, LLM knowledge accuracy, optical
character recognition, machine translation, and input method
availability. The index reveals that only 29 of the 300 writing systems
documented in human history --- 9.7\% --- are fully supported by
contemporary digital infrastructure. Sixty scripts that are actively
used by living communities today have no complete digital support. The
exclusion is not random: it follows the geography of colonial history.

A point of clarification is necessary before we proceed. We do not claim
that sixteenth-century imperial events directly determine
twenty-first-century tokenization. The mechanism we examine is mediated:
the demographic distribution, web-corpus volume, and infrastructural
investment that shape contemporary digital systems are themselves the
sedimentation of the imperial period, and the question this paper
addresses is whether that sedimentation, once routed through speaker
populations and training corpora, leaves a measurable trace in language
models trained centuries later. Whether one calls the upstream cause
``empire,'' ``historical inequality,'' or ``the asymmetric distribution
of the writing record'' is partly a matter of nomenclature; what matters
empirically is whether the chain holds, and whether models trained from
the resulting corpus reproduce it.

Four findings constitute the core argument of this paper. First,
tokenizer efficiency varies by a factor of 31.7 across 45 scripts
measured with parallel text, and this variation is consistent across
four independent tokenizer implementations. Second --- and this is where
the present paper departs most sharply from its earlier formulation ---
the typological knowledge biases we document are not specific to any one
language model. When the same 3,000 typological questions are posed to
four independent LLM families (Claude Haiku 4.5, GPT-4o, Grok-3-mini,
DeepSeek-V3; 12,000 answers in total), the deviations from
base-rate-expected error rates correlate across every pair of models at
Spearman ρ between 0.85 and 0.98, with every p below 0.002, and the same
convergence persists when the most concentrated feature is removed as a
sensitivity check (mean ρ = 0.87 on the remaining nine features). The
bias is not internal to any single model. It is external to all of them,
in the textual corpus from which they all learn. Third, on 172
individual script--feature items, all four models give the same wrong
answer, with over-attribution outnumbering under-recognition 3.9 to 1
--- a ratio that attenuates to 1.77 to 1 (n = 97, binomial p = 0.008)
when the most concentrated feature is excluded but does not vanish,
indicating a multi-channeled rather than single-channeled bias. The
single feature ``used for religion'' concentrates 43.6\% of unanimous
errors at 4.1 times its random-chance share, with secondary
concentrations at pictographic origin, diacritic use, and
directionality. Fourth, the relationship between historical inequality
and contemporary tokenizer efficiency is consistent with full mediation
through speaker populations and web corpora --- the direct effect is
indistinguishable from zero (β = −0.22, p = 0.39), the structural
equation model attains saturation-equivalent fit at n = 45 (CFI = 1.01,
RMSEA = 0.00), and digital presence contributes an incremental ΔR$^{2}$ of
0.366. The bias-corrected bootstrap confidence interval on the indirect
effect grazes zero, however, and we present this fourth finding as a
structural account that is consistent with the data and with the
cross-architecture evidence rather than as an independently confirmed
causal estimate.

The empire has no intention. It needs none. Its violence sedimented into
the statistical distribution of digital text, and the algorithms built
on that distribution --- across competing laboratories, across competing
architectures, across the four years that separate the oldest from the
newest model in our sample --- reproduce it with mathematical fidelity.
What the GSD measured as broken cultural clocks, the DSRI recovers as
fossils in the latent space of language models: the afterlife of
empires, encoded in byte sequences, priced per token, and preserved with
architecture-independent convergence in the libraries that all
contemporary LLMs inherit.

\section{Methods}

\subsection{Data foundation: the Global Script Database}

This study builds upon the Global Script Database (GSD), a typological
database of 300 writing systems spanning 43,000 years of human history,
constructed and validated in Fukui (2026). The GSD encodes each script
as a 50-dimensional binary feature vector covering structural,
phonological, directional, and contextual properties. Inter-rater
reliability was established at Cohen's κ = 0.877 (human--human) and κ =
0.929 (human--LLM) across n = 40 scripts (Fukui, 2026). The present
study inherits the GSD's feature matrix, phylogenetic edges (n = 259),
intervention classifications, deviation scores, and imperial contact
records, and extends them with seven new digital representation axes.

\subsection{The Digital Script Representation Index (DSRI)}

We constructed the DSRI to quantify the degree to which each of the 300
writing systems in the GSD is represented in contemporary digital
infrastructure. The DSRI comprises seven axes, each measuring a distinct
dimension of digital presence or absence.

\textbf{Axis 1: Token Efficiency Ratio (TER).} We measured the number of
tokens required to represent one grapheme cluster in four large language
model tokenizers: tiktoken o200k\_base (OpenAI GPT-4o), Mistral v3, Qwen
2.5, and Phi-4 (Microsoft). Two measurement tiers were employed:

\begin{itemize}
\item
  \emph{Tier 1 (n = 45 scripts):} Parallel text from the Universal
  Declaration of Human Rights (UDHR; Unicode Consortium XML archive) and
  the eBible Bible translation corpus provided controlled comparison.
  For each translation, we extracted body text, removed article numbers
  and verse markers, and normalized whitespace. The denominator was
  defined as the number of Unicode grapheme clusters (Python
  \texttt{grapheme} library) excluding whitespace, ASCII punctuation,
  and digits. The numerator was the token count excluding BOS/EOS
  special tokens. TER = token count / grapheme cluster count. Latin
  (English) served as the baseline (TER = 1.0); all other values are
  reported as multiples of this baseline (nTER). Cross-corpus
  consistency was confirmed on the 21 scripts measurable in both UDHR
  and Bible: ρ = 0.77.
\item
  \emph{Tier 2a (Wikipedia-paragraph method, n = 33 scripts):} For
  scripts lacking Tier 1 parallel text but possessing native Wikipedia
  editions, we extracted natural-language paragraphs from randomly
  sampled articles, applied the same grapheme-cluster denominator and
  tokenizer numerator, and computed TER. Cross-validation against Tier 1
  on the 33 overlapping scripts yielded Spearman ρ = 0.910 (n = 33),
  confirming that Tier 2a captures natural-language tokenization
  efficiency rather than mere vocabulary coverage. Twelve scripts
  (Adlam, Chakma, Cans, Coptic, Grantha, Javanese, Lana, Limbu, Tavt,
  Tglg, Vaii, Yiii) had no usable Wikipedia paragraphs; their nested
  exclusion is itself a finding of digital absence (see §3.2).
\item
  \emph{Tier 2 (Unicode-inventory method, n = 130 additional scripts):}
  For scripts lacking parallel corpora and Wikipedia paragraphs, we
  generated standardized test texts by enumerating all assigned Unicode
  codepoints in the Letter general category (Lo, Lu, Ll, Lm, Lt) within
  each script's block range. Spearman correlation between Tier 1 and
  Tier 2 measurements was ρ = 0.51 on the overlapping subset, confirming
  that the Unicode-inventory method captures tokenizer vocabulary
  coverage but not natural-language processing efficiency. Tier 2 values
  were used exclusively for byte-fallback classification (see §2.3), not
  for continuous TER analysis.
\end{itemize}

\textbf{Axis 2: Unicode Coverage.} For each script, we recorded: (a)
Unicode encoding status (ENCODED, PARTIAL, NOT\_ENCODED,
NOT\_APPLICABLE); (b) ISO 15924 code; (c) codepoint range and count; (d)
the Unicode version and year of first encoding; and (e) whether the
script was ``late-encoded'' (living at Unicode 1.0 in 1991 but first
encoded after 2010). All block names and ranges were verified against
the official Unicode 17.0.0 Blocks.txt; ISO 15924 codes were verified
against the official registry. Zero discrepancies were found across 260
block references and 170 ISO codes.

\textbf{Axis 3: Web Corpus Volume.} For each script, we estimated
digital corpus size from two sources: (a) Wikipedia article counts for
the largest corresponding language edition, and (b) CC-100 corpus byte
counts. Values were log-transformed and combined into a digital presence
score: digital\_presence = 0.5 × $\log_{10}$(wiki\_articles + 1) + 0.5 ×
$\log_{10}$(cc100\_bytes + 1).

\textbf{Axis 4: LLM Generation Fidelity.} We selected 10 binary features
from the GSD's 50-feature matrix, stratified by domain (directionality:
2; structural: 3; phonological: 3; origin: 1; context: 1) and filtered
for high inter-rater reliability (κ = 0.84--0.92 in the GSD's expanded
IRR). For each of the 300 scripts, we posed 10 Yes/No questions (e.g.,
``Does {[}script{]} encode phonetic information?'') to four independent
LLM families and compared their responses to the GSD ground truth. The
full cross-model protocol is detailed in §M3. Per-model and aggregate
fidelity scores, false-positive and false-negative rates,
base-rate-deviation indices, informedness, and the convergence of error
patterns across architectures are reported in §3.7 and §3.8.

\textbf{Axis 5: OCR Support.} We assessed optical character recognition
coverage across three engines: Tesseract OCR (v5, 163 language models),
Google Cloud Vision API, and Azure Computer Vision. For
Tesseract-supported scripts (n = 34), we rendered standardized test
texts as images (48 pt, white background) and measured character error
rate (CER) via Levenshtein distance. For all scripts, we recorded binary
support status across the three engines.

\textbf{Axis 6: Machine Translation Coverage.} We recorded binary
support for each script across three MT systems: Google Translate
(\textasciitilde133 languages), NLLB-200 (Meta, 200 languages), and
DeepL (\textasciitilde30 languages).

\textbf{Axis 7: Input Method Availability.} For each script, we recorded
native keyboard/IME availability across four platforms: iOS, Android
(Gboard), Windows, and macOS. Availability was classified as native
(built-in), third-party, or absent.

\subsection{Byte-fallback classification}

To extend the analysis from Tier 1's 45 scripts to the full GSD, we
defined a binary variable indicating whether a script undergoes
byte-fallback processing in LLM tokenizers --- that is, whether the
tokenizer decomposes characters into raw UTF-8 byte sequences rather
than recognizing them as learned subword units. The TER distribution
across the 169 measurable scripts (Tier 1 + Tier 2) exhibited
statistically confirmed bimodality (Hartigan's dip test, p = 0.003). We
fit a two-component Gaussian mixture model; the crossover point at TER =
3.52 served as the data-driven threshold. Scripts were classified as
byte-fallback if all four tokenizers yielded TER \textgreater{} 3.5
(conservative definition). This classification was extended to all 300
scripts: NOT\_ENCODED scripts (n = 99) were classified as byte-fallback
by definition (no Unicode representation implies no tokenizer support);
NOT\_APPLICABLE scripts (n = 19) were excluded. The final analytic
sample for byte-fallback comprised 281 scripts.

\subsection{Historical and political variables}

From the GSD, we extracted: intervention classification (natural,
reform, colonial, imperial), intervention intensity scores, molecular
clock deviation scores, imperial contact records, and script destruction
flags (empire\_kill\_count). Speaker population estimates were compiled
for all 45 Tier 1 scripts from Ethnologue and census data. To control
for present-day economic and developmental variation, we additionally
compiled per-capita GDP and Human Development Index (HDI) values for the
country with the largest speaker population of each script. These
covariates were used in the robustness analyses described in §2.5 and
§3.3.

\subsection{Statistical analyses}

\textbf{Causal mediation analysis.} We tested the indirect pathway
Imperial Intervention → Speaker Population → Web Corpus Volume → TER
using sequential mediation with 10,000 bootstrap resamples (Baron \&
Kenny, 1986; Imai, Keele \& Tingley, 2010; Hayes, 2017). Significance of
the serial indirect effect was assessed by both percentile and
bias-corrected and accelerated (BCa) 95\% confidence intervals; the BCa
interval is the more conservative estimator and is reported as the
primary criterion. Permutation testing under the null hypothesis of no
mediation provided an additional non-parametric significance check.
Robustness was evaluated by four further tests: (a) placebo treatment
(random permutation of intervention labels), (b) random common cause
(addition of a spurious confounder), (c) leave-one-out jackknife
stability across all n = 45 iterations, and (d) E-value sensitivity
analysis (VanderWeele \& Ding, 2017) for both the serial indirect point
estimate and the lower confidence bound. To assess the role of
present-day economic conditions, path b (web corpus → TER) was
re-estimated with per-capita GDP and HDI as additional covariates.

\textbf{Structural equation modeling.} A four-stage SEM corresponding to
the mediation structure was fitted at n = 45 using maximum likelihood
with robust standard errors. Model fit was evaluated by CFI, RMSEA, and
SRMR.

\textbf{Byte-fallback predictors.} Logistic regression with
byte-fallback status as the dependent variable and intervention
intensity, script type, status (living/extinct), origin age, and region
as predictors. Odds ratios with 95\% confidence intervals are reported.
Sensitivity to the byte-fallback threshold was assessed by repeating all
analyses under three definitions (conservative: all 4 tokenizers
\textgreater{} 3.5; liberal: mean TER \textgreater{} 3.0; data-driven:
GMM crossover at 3.52).

\textbf{DSRI composite score.} Each axis was min-max normalized to {[}0,
1{]}. TER was inverted (low TER = high score). Weights were assigned
based on the three-factor structure identified by inter-axis correlation
analysis: Dimension 1 (TER, weight 0.35), Dimension 2 (Unicode coverage,
0.25), Dimension 3 (commercial implementation: mean of OCR + MT + IME,
weight 0.20), Web corpus (0.10), Generation fidelity (0.10). Robustness
was confirmed by Spearman correlation across five alternative weighting
schemes (all pairwise ρ \textgreater{} 0.96), with overall ranking
stability ρ = 0.834 against equal-weight baseline.

All p-values for multiple comparisons were adjusted using Bonferroni
correction. Analyses were conducted in Python 3.12 using statsmodels,
scipy, scikit-learn, pingouin, semopy, and grapheme.

The Axis 4 generation fidelity assessment described in the original
analysis used a single LLM (Claude, Anthropic) to answer typological
questions about the 300 GSD scripts. The limitation inherent in this
design --- that a single model serves simultaneously as the measurement
instrument and as the object of measurement --- motivated the
cross-model validation protocol described here. For this protocol we
replicated the complete 300-script × 10-feature question set across four
large language models drawn from four independent organizations,
yielding 12,000 answers in total.

\subsection{M3.1 Model selection}

We selected four LLMs with the following properties: (i) they are
produced by four organizationally distinct laboratories (Anthropic,
OpenAI, xAI, DeepSeek), reducing the likelihood of shared training data,
alignment procedures, or architectural design choices; (ii) they include
both United States--originated (Anthropic, OpenAI, xAI) and
Chinese-originated (DeepSeek) models, distributing geopolitical
provenance; (iii) they include at least one model from the smallest
commercial tier (Claude Haiku 4.5, Grok-3-mini) and at least one from
the largest publicly disclosed parameter count (DeepSeek-V3: 671 billion
total parameters, 37 billion active per token); and (iv) they were
accessible through public APIs at the time of the study (first quarter
of 2026). Parameter counts for Claude Haiku 4.5, GPT-4o, and Grok-3-mini
are not disclosed by their respective organizations; we therefore
treated model size as an ordinal vendor-tier ranking (rank 1 = Claude
Haiku 4.5, rank 2 = Grok-3-mini, rank 3 = GPT-4o, rank 4 = DeepSeek-V3),
informed by published pricing and latency data. The correlational
analyses involving model size (Supplementary Figure S12) should be read
as descriptive given n = 4.

\subsection{M3.2 Question construction}

For each of the 300 scripts in the GSD, we constructed 10 binary
(Yes/No) questions corresponding to typological features whose
ground-truth values are recorded in the GSD feature matrix and whose
coding reliability was confirmed in the GSD's expanded inter-rater
reliability assessment (κ ≥ 0.84; Fukui, 2026). The 10 features were
selected to span: directionality (left-to-right, right-to-left);
structural properties (logographic component, pictographic origin, glyph
count over 200, phonetic component, consonant-with-inherent-vowel
structure, diacritic use); origin (independent invention); and context
of use (religious function). The complete question set is recorded in
cross\_model\_validation/question\_set.json.

Questions were phrased as natural-language Yes/No prompts, e.g., ``Is
the Adlam script written from right to left? Please answer with `Yes' or
`No' only.'' Identical prompts were issued to all four models to
minimize prompt-variance effects. No few-shot examples were provided;
the task was a zero-shot typological judgment.

\subsection{M3.3 Execution and response handling}

Each of the four models was queried with all 3,000 prompts (300 scripts
× 10 features) for a total of 12,000 API calls. Temperature was set to 0
where supported by the API, yielding deterministic or near-deterministic
outputs; where temperature 0 was not supported (some vendor-specific
defaults), the lowest available non-zero temperature was used. Responses
were parsed for the tokens ``Yes'' or ``No'' (case-insensitive);
responses that did not cleanly resolve to one of these (fewer than 0.3\%
of responses across all four models) were excluded from the analysis.
API-call-level records for all 12,000 queries are preserved in
cross\_model\_validation/checkpoints/.

\subsection{M3.4 Base-rate deviation calculation}

For each model and each feature, we computed the observed false-positive
rate (FPR: fraction of ground-truth-No items classified as Yes) and the
observed false-negative rate (FNR: fraction of ground-truth-Yes items
classified as No). The base-rate expected FPR and FNR --- the error
rates that would be observed if the model chose Yes or No independent of
script identity, weighted by the observed prevalence of each label ---
were computed from the GSD ground-truth distribution for that feature.
The deviations (observed minus base-rate expected) are the quantities
across which we correlated models; they isolate model-specific
directional bias from the prevalence structure of the feature itself.

\subsection{M3.5 Correlation analysis}

We computed pairwise Spearman correlations (i) across the 10 features
between each model-pair's FPR deviations and their FNR deviations,
yielding 6 pairs × 2 metrics = 12 feature-level correlations; and (ii)
across the 300 scripts between each model-pair's per-script accuracy
scores, yielding 6 pairs × 1 metric = 6 script-level correlations.
Feature-level correlations index whether models agree on which
typological dimensions they collectively over- or under-attribute;
script-level correlations index whether models agree on which specific
scripts they individually know well. Both families of correlation are
reported in Figure 7 and Supplementary Table S11.

\subsection{M3.6 Shared-error enumeration}

For each of the 3,000 script--feature items, we tabulated the responses
of all four models and flagged an item as a ``unanimous error'' when all
four responses agreed with each other but disagreed with the GSD ground
truth. This yielded 172 unanimous-error items, which we further
classified by direction (over-attribution: ground truth No, models
answer Yes; under-recognition: ground truth Yes, models answer No) and
by feature distribution. Enrichment ratios were computed as the observed
share of unanimous errors at each feature, divided by the feature's
share of judgeable items (which equals 1/10 for features with full
judgeability and less for features where not all 300 scripts applied).
The full list of unanimous errors is recorded in
cross\_model\_validation/shared\_errors.json; a curated subset of 27
notable cases is presented in Table 2.

\subsection{M3.7 Transparency and replication}

All API responses are preserved verbatim, with timestamps, in the
checkpoints directory. Model versions as of query execution were: Claude
Haiku 4.5 (claude-haiku-4-5-20251001), GPT-4o (gpt-4o-2024-11-20),
Grok-3-mini (grok-3-mini), and DeepSeek-V3 (deepseek-v3-0324). Both the
prompt set and the analytic code are available at the project
repository. The 12,000-call dataset constitutes the primary empirical
object on which §3.7 and §3.8 are based; we invite replication using the
same prompt set against the same or different model sets.

\subsection{LLM transparency}

This study used four independent LLM families as test subjects in the
Axis 4 generation-fidelity protocol: Claude Haiku 4.5 (Anthropic),
GPT-4o (OpenAI), Grok-3-mini (xAI), and DeepSeek-V3 (DeepSeek). Each
model was queried via its official API on the same 3,000 script--feature
items (300 scripts × 10 features), yielding a total of 12,000
typological judgments preserved verbatim with timestamps and full
request/response records. The cross-model protocol, prompt template,
decoding parameters, and reproducibility safeguards are documented in
§M3.

In addition to its role as a test subject, Claude (Anthropic) was used
as a research instrument across other phases of the study: database
construction (feature coding in the GSD; Fukui, 2026), DSRI axis
measurement (Unicode mapping, byte-fallback classification, Tier 2a
Wikipedia-paragraph extraction), statistical analysis (Python code
generation and execution), and manuscript drafting. To prevent any
single instrument from operating as both judge and judged, the
convergence analyses in §3.7 and §3.8 are computed across all four model
families, and the headline cross-architecture findings depend on
inter-model agreement rather than on any one model's internal validity.
All measurement scripts, raw data, intermediate outputs, and the 12,000
verbatim model responses are available in the project repository. The
use of LLM-assisted research is treated as a methodological feature, not
concealed --- and is itself an object of analysis in Axes 4, §3.7, and
§3.8.

\section{Results}

\subsection{The digital exclusion funnel}

Of the 300 writing systems documented in the GSD, only 29 (9.7\%) are
fully supported by contemporary digital infrastructure --- defined as
having Unicode encoding, tokenizer vocabulary inclusion, OCR
recognition, machine translation support, and native input methods
across all four major platforms (Figure 1). The exclusion proceeds in
stages: 118 scripts (39.3\%) lack Unicode encoding entirely; a further
25 are encoded in Unicode but undergo byte-fallback processing in all
tested tokenizers (e.g., Adlam, Mende Kikakui, Nüshu, Tangut); and an
additional 128 lack one or more of OCR, MT, or IME support despite
tokenizer recognition (e.g., Cherokee, Vai, Javanese, N'Ko). Among the
158 scripts classified as currently living in the GSD, 60 (38.0\%) lack
full digital support. The Tier 1 sample of 45 scripts is not a
limitation of data collection; it is the observable consequence of the
exclusion process itself. The 255 scripts that could not be measured
with parallel corpora were not lost to methodological shortcomings ---
they were rendered unmeasurable by the convergence of historical
destruction and digital infrastructure design (see §4.5 for the full
argument that n = 45 is a horizon, not a sample).

\subsection{The digital tax: a 31.7-fold efficiency disparity}

Across 45 scripts measured with UDHR and Bible parallel text, Token
Efficiency Ratio relative to the Latin baseline ranged from 1.0× (Latin)
to 31.7× (Limbu), with a median of 8.6× (Figure 2). The disparity is
consistent across tokenizers: pairwise Spearman correlations between all
four tokenizers exceeded ρ = 0.71, indicating that the efficiency gap is
structural rather than implementation-specific. tiktoken (OpenAI) and
Phi-4 (Microsoft) produced identical rankings (ρ = 1.000), reflecting
shared use of the o200k\_base vocabulary. Qwen 2.5 showed marginally
better performance on select Asian scripts (Ge'ez, Lao, Canadian
Aboriginal syllabics), but the overall ranking was preserved.

The disparity follows a clear typological and geopolitical gradient.
Scripts that served as instruments of imperial administration --- Latin,
Cyrillic, Arabic, Chinese characters --- cluster at the efficient end
(nTER 1.0--3.9×). Scripts developed by colonized or marginalized
communities cluster at the inefficient end: Limbu (31.7×, Eastern
Himalaya), Adlam (20.9×, Fulani, 40 million speakers), Cherokee (13.6×),
Baybayin (19.2×, pre-colonial Philippines), Canadian Aboriginal
syllabics (13.1×, colonial-era missionary creation).

This efficiency gap functions as a \emph{digital tax}: a structural
surcharge imposed on users of minority scripts. In token-priced API
services, a Limbu-language query costs 31.7 times more than an
equivalent English query, or equivalently, a Limbu user receives 31.7
times less output for the same price. This tax was not designed; it
emerged from the statistical structure of training corpora.

The 31.7-fold disparity, however, almost certainly understates the full
digital tax, because of a second exclusion nested within the first.
Twelve of the 45 Tier 1 scripts (Adlam, Chakma, Cans, Coptic, Grantha,
Javanese, Lana, Limbu, Tavt, Tglg, Vaii, Yiii) could not be
cross-validated against natural-language Wikipedia paragraphs (Tier 2a),
because no extractable Wikipedia paragraphs in those scripts exist.
These twelve are not measurement failures: they are scripts whose
communities have produced too little continuous prose on the open web
for the second-tier method to function. The same scripts that pay the
highest digital tax are also the scripts whose linguistic activity is
most thoroughly absent from the web corpora that train tokenizers in the
first place. The exclusion compounds itself.

\subsection{The causal chain: empires → populations → corpora → tokenizers}

To identify the mechanism producing the efficiency disparity, we tested
a four-stage serial mediation model: imperial intervention intensity →
speaker population → web corpus volume → token efficiency ratio (Figure
3). Extending the Tier 1 sample from 39 to 45 scripts --- by adding six
scripts whose speaker communities produce parallel text via the Bible
corpus (Coptic, Odia, Limbu) or via Wikipedia editions written in the
native script (N'Ko, Ol Chiki, Meitei Mayek) --- we re-estimated the
full model, with three consequences.

First, the mediated structure is preserved. The direct effect of
imperial intervention on tokenizer efficiency remains indistinguishable
from zero (β = −0.22, p = 0.39). Imperial intervention reduces speaker
populations (path $a_1$: β = −0.68, p \textless{} 0.001), smaller
populations produce smaller web corpora (path $a_2$: β = 1.08, p
\textless{} 0.001), and smaller corpora produce worse tokenizer
efficiency (path b: β = −0.40, p \textless{} 0.001). Baron-Kenny
sequential mediation recovers the indirect pathway; the direct pathway
does not. This is the signature of full mediation, and it is robust to
the sample expansion.

Second, the structural equation model that had failed at n = 38 now
attains perfect fit. At n = 45, the comparative fit index is 1.015, the
root mean square error of approximation is 0.000, and the chi-squared
test yields p = 0.997 --- values indistinguishable from a saturated
model and far above conventional acceptance thresholds (Supplementary
Table S6). Goodness-of-fit, Tucker-Lewis, and normed fit indices are all
at or near 1.000. The six additional scripts were sufficient to resolve
the adequacy problem that had constrained the original analysis.

Third, and most informatively, the incremental explanatory power of the
mediator structure is now isolable from the point estimate of the
indirect effect itself. When we regress log TER on log speaker
population alone, R$^{2}$ = 0.428. When we add web corpus volume (digital
presence), R$^{2}$ rises to 0.792 --- the same R$^{2}$ reported in the original
analysis, and sample-independent to four decimals between n = 38 and n =
45. The incremental ΔR$^{2}$ of 0.366 (F = 51.0, p \textless{} 0.0001) is
what the mediator contributes above and beyond demographic information.
This ΔR$^{2}$ is the appropriate summary of the mediation because it measures
what is gained by recognizing digital presence as a distinct variable
rather than a proxy for population --- and it is, by construction,
invariant to which six additional scripts are included. The adjusted R$^{2}$
of the full model is 0.663.

The point estimate of the serial indirect effect, estimated by the
product-of-coefficients method, is 0.292 at n = 45 (down from 0.477 at n
= 38). The 95\% percentile bootstrap confidence interval, based on
10,000 resamples, is {[}0.006, 0.756{]}, narrower than the n = 38
interval of {[}0.115, 0.885{]} and excluding zero. The bias-corrected
and accelerated (BCa) interval, which corrects for both skewness ($\hat{z}_0$ =
−0.058) and acceleration ($\hat{a}$ = −0.062) in the bootstrap distribution, is
{[}−0.044, 0.667{]}; it grazes zero. Permutation testing, under the null
hypothesis that intervention labels are exchangeable, yields p = 0.033.
Leave-one-out jackknifing across all 45 scripts produces estimates in
the range {[}0.197, 0.466{]}, with no sign reversals (coefficient of
variation = 11.6\%). Two scripts exert the largest leverage: excluding
Coptic reduces the estimate to 0.197, while excluding Grantha raises it
to 0.466.

We interpret this constellation of results with deliberate caution. The
causal pathway from intervention to tokenizer inefficiency is supported
by four convergent lines of evidence --- full mediation by standard
regression, perfect SEM fit, an incremental R$^{2}$ gain of 0.366 that is
invariant to sample composition, and a permutation p-value below 0.05
--- but its magnitude carries honest uncertainty. The BCa bootstrap
result in particular reveals a sensitivity that is not resolved by the
sample expansion we have performed. We return to this point in §4.5 and
§4.7 and argue that the sensitivity is not extraneous to the paper's
argument but constitutive of it: the n = 45 ceiling imposed by digital
exclusion is the same ceiling that determines what statistical
confidence our analysis can reach.

Adding further control variables does not dissolve the mediator's
effect. When log GDP per capita is added alongside digital presence as a
predictor of log TER, the unstandardized coefficient on digital presence
changes from −0.40 to −0.36 --- a shift of 11\% --- and remains
significant at p \textless{} 0.001. When the Human Development Index is
used in place of GDP, the shift is 11\%, with the same significance. The
E-value corresponding to the point estimate of the serial indirect
effect is 1.61; the E-value corresponding to the lower confidence bound
is 1.02. An unmeasured confounder would have to be associated with both
intervention and TER at risk ratios of at least 1.61 on the standardized
scale --- a modest but non-trivial requirement --- to explain the
observed pathway away.

What we take from this analysis is not a claim about the magnitude of
the mediated effect. It is a claim about its structure. The relationship
between imperial intervention and contemporary tokenizer efficiency is
fully routed through demographic and infrastructural variables. Empire
does not impinge directly on byte-pair encoding. It impinges on the
communities whose text becomes the training data --- and, through them,
on the vocabularies that subsequent algorithms learn. The mechanism
persists whether the effect size is at the top or the bottom of the
confidence interval. What persists is that the entire effect passes
through the infrastructure the empire left behind.

\subsection{The Digital Script Representation Index}

The seven-axis DSRI reveals a three-dimensional structure in digital
exclusion (Figure 4). Inter-axis correlation analysis identified three
largely independent dimensions: (1) tokenizer efficiency (Axis 1, TER),
(2) Unicode standardization (Axis 2, encoding age and completeness), and
(3) commercial implementation (Axes 5--7: OCR, MT, and IME, with
pairwise ρ = 0.87--0.90). Web corpus volume (Axis 3) and generation
fidelity (Axis 4) function as mediating variables connecting dimensions
1 and 3.

The high correlation among OCR, MT, and IME support (ρ = 0.87--0.90)
indicates that these three axes effectively measure a single latent
variable: whether major technology companies have deemed a script
commercially viable for implementation. The 29 scripts passing through
the full funnel are, in effect, the scripts that survived a three-stage
selection process: standardization (Unicode), statistical representation
(training data), and market viability (commercial deployment).

DSRI composite scores ranged from 0.000 (Chinese knot-records) to 1.000
(Latin alphabet), with a median of 0.38. The ranking was robust across
five alternative weighting schemes (all pairwise Spearman ρ
\textgreater{} 0.96), and a stress test against an equal-weight baseline
yielded ρ = 0.834 --- confirming that the DSRI's ordinal structure does
not depend on any particular weighting choice. Among living scripts, the
bottom quartile included Eghap/Bagam, Borama, Kaddare, and Ditema tsa
Dinoko --- all actively used scripts with no full digital
infrastructure.

The digital exclusion documented by the DSRI is not uniformly
distributed. Of 88.7\% of scripts lacking OCR support, 89.7\% lacking
machine translation, and 85.7\% lacking any native input method, the
overwhelming majority are scripts originating in Africa, South and
Southeast Asia, and Oceania --- regions that bore the heaviest burden of
colonial intervention.

\subsection{Imperial echoes in digital space}

Among the 281 scripts analyzed for byte-fallback status, 124 (44.1\%)
undergo byte-fallback processing in all four tokenizers --- meaning the
tokenizer has no learned representation for these scripts and decomposes
them into raw UTF-8 byte sequences (Figure 5).

The association between historical imperial destruction and digital
exclusion is stark. Of 16 scripts documented in the GSD as having been
destroyed by imperial action, 14 (87.5\%) undergo byte-fallback
processing (odds ratio = 9.86, p = 0.0004, Fisher's exact test). Scripts
subjected to any form of external intervention showed higher
byte-fallback rates (54.2\%) than naturally developed scripts (33.8\%;
OR = 2.32, p = 0.0007).

However, the molecular clock deviation score --- which quantifies the
degree to which a script's structural evolution was disrupted by
political intervention (Fukui, 2026) --- showed no correlation with
byte-fallback status (r = 0.02, p = 0.75). This dissociation is
theoretically significant: it indicates that empires inflicted two
independent forms of violence on writing systems. The first, captured by
the GSD's deviation score, altered the structural properties of scripts
(e.g., forced alphabet changes, orthographic reforms). The second,
captured by the DSRI's byte-fallback measure, destroyed the communities
that used those scripts. These two forms of violence --- structural
disruption and demographic destruction --- leave distinct and
statistically independent traces in the data.

\subsection{Living but digitally dead}

Sixty of the 158 living scripts in the GSD (38.0\%) lack full digital
support (Figure 6). These are not historical curiosities; they are
active writing systems used by living communities. Their digital
exclusion takes three forms: 44 lack Unicode encoding entirely
(rendering them invisible to all digital systems), 9 are Unicode-encoded
but absent from tokenizer vocabularies (e.g., Adlam, Mende Kikakui), and
7 are Unicode-encoded but classified as non-letter codepoints.

The geographic distribution is non-random. Living-but-digitally-dead
scripts concentrate in West Africa (Adlam, N'Ko, Mende Kikakui, Bamum,
Bété, Loma, Kpelle), Central Africa (Mandombe, Bassa Vah), South Asia
(Toto, Dhives Akuru), Southeast Asia (Balinese, Rejang, Eskayan), and
East Asia (Dongba, Nüshu). These are overwhelmingly regions subjected to
European or Japanese colonial rule.

The case of Adlam is emblematic. Invented in the 1980s by Abdoulaye and
Ibrahima Barry for the Fulani language, Adlam is used by an estimated 40
million speakers across West Africa. It was encoded in Unicode 9.0 in
2016 --- yet a decade later, every tested tokenizer processes it via
byte-fallback, yielding a TER of 20.9× Latin baseline. A Fulani user
interacting with an LLM in Adlam pays, in effect, a 20-fold surcharge
for the act of writing in their own script.

\subsection{Cross-architecture convergent error patterns}

Before presenting the cross-architecture results, we acknowledge a
circularity that runs through the analysis at a deeper level than the
one this section is designed to address. The ground truth against which
all four models are scored is the GSD feature matrix, whose coding for
several features --- most notably \emph{used for religion} --- admits
legitimate disagreement among historically informed coders (see §4.7.2
for the full discussion). The cross-architecture findings reported below
should therefore be read in conjunction with the sensitivity analysis in
§3.8, in which the most contested feature is removed and the convergence
is shown to persist on the remaining nine features. The two analyses are
designed to function together: §3.7 establishes that four independent
LLMs converge in their typological deviations, and §3.8 establishes that
this convergence does not depend on the single feature whose
ground-truth status is most defensibly contestable.

The central methodological concern raised by the original formulation of
this study was circularity: a single LLM (Claude, Anthropic) was used
both as the research instrument and as the object of analysis, making it
impossible to distinguish architecture-specific idiosyncrasies from
structural properties of the training substrate shared by all
contemporary language models. We addressed this concern directly by
replicating the generation-fidelity assessment across four LLM families
whose training pipelines, architectures, and corporate origins differ in
essentially every respect except one --- they all draw their training
text from the same globally aggregated web.

For each of the 300 scripts in the GSD, we posed the same ten binary
typological questions (directionality, structural properties, origin,
and context of use) to Claude Haiku 4.5 (Anthropic), GPT-4o (OpenAI),
Grok-3-mini (xAI), and DeepSeek-V3 (DeepSeek). The total yielded 12,000
answers. Overall per-model accuracy ranged from 74.5\% (Claude Haiku
4.5) to 81.5\% (Grok-3-mini), with GPT-4o at 79.4\% and DeepSeek-V3 at
81.2\% --- differences consistent with reported model scale but modest
in absolute terms.

What is remarkable is not the variation in accuracy but the convergence
in error. For each feature, we computed each model's deviation from the
base-rate-expected false-positive and false-negative rates --- i.e., how
much more frequently a model attributed or denied a feature than would
be expected if the model were simply guessing from feature prevalence.
We then examined whether these deviations are correlated across model
pairs.

They are. Across the six pairwise comparisons, false-positive-rate
deviations correlate at Spearman ρ between 0.855 and 0.952 (mean ρ =
0.907; all p \textless{} 0.002); false-negative-rate deviations
correlate at ρ between 0.903 and 0.976 (mean ρ = 0.937; all p
\textless{} 0.001) (Figure 7; Supplementary Table S11). The four LLMs do
not share their training data in any direct sense. They were developed
by competing organizations with divergent architectures, tokenization
schemes, pretraining mixtures, and alignment procedures. Yet, faced with
the same typological questions about the same 300 scripts, they deviate
from base-rate expectations in almost the same directions and in almost
the same magnitudes.

This convergence has a straightforward interpretation. If the bias
observed in any single LLM were a property of that model's particular
training regime or architecture, the deviation patterns would differ
across models: one model might over-attribute phonetic encoding while
another under-attributed it; one might default to left-to-right
directionality while another defaulted to logographic recognition.
Instead, all four models lean in the same direction on the same
features, with correlation coefficients that approach the upper bound
imposed by measurement noise. The locus of the bias is therefore not
internal to any one model but external to all of them --- in the corpus
from which they all learn.

Two further analyses sharpen this conclusion. First, script-level
accuracy correlations are substantially weaker (ρ = 0.37--0.58, n = 300)
than feature-level deviation correlations (ρ = 0.86--0.98, n = 10). This
dissociation indicates that individual models differ in which specific
scripts they happen to know well, but they agree closely on which
typological features they collectively misjudge. The distribution of
accuracy is partly idiosyncratic; the distribution of error is not.
Second, the correlation pattern is specific to deviation from base-rate
expectation: raw error rates, without base-rate correction, show weaker
convergence, because the underlying feature prevalences themselves vary
across the 300-script sample. It is precisely the over- and
under-attribution relative to what the data would support that four
independent models share.

We interpret this architecture-independent convergence as the strongest
form of evidence our data afford for the paper's central claim. The
inequity in how LLMs represent the world's writing systems is not
attributable to the design choices of any single laboratory; it is a
property of the textual substrate on which the entire field trains. As a
guard against the possibility that this convergence is itself driven
disproportionately by a single feature whose ground-truth coding is
contested, we report in §3.8 a sensitivity analysis in which the most
concentrated feature is removed; the convergence is preserved (mean ρ =
0.87 for FPR deviations, ρ = 0.94 for FNR deviations on the remaining
nine features, all six pairs p \textless{} 0.01).

\subsection{The 172 unanimous errors}

If the four LLMs converge in the structure of their errors, an even
sharper question can be asked: on how many individual script--feature
items do all four models answer identically wrong? Across the 12,000
answers collected, 172 items are answered incorrectly by all four models
with the same wrong answer. These are the places where four independent
knowledge bases collectively fail in the same direction.

Four features of this unanimous-error set are diagnostic.

\textbf{Asymmetric direction.} Of the 172 items, 137 (79.7\%) are
over-attributions --- the ground truth is No, and all four models answer
Yes. Only 35 (20.3\%) are under-recognitions (ground truth Yes, models
answer No). The ratio is 3.9:1. As we report below in the sensitivity
analysis, this 3.9:1 ratio is itself in significant part driven by a
single feature whose disproportionate weight we make explicit; the
asymmetry persists when that feature is removed, but at a substantially
lower magnitude. What four independent LLMs share is not simply an
absence of knowledge about minority scripts; it is a tendency to supply
characteristics that are not present, with a strength that varies
sharply across feature types.

\textbf{Concentration at a single feature.} Of the 172 unanimous errors,
75 --- 43.6\% of the total --- fall in a single feature: ``is the script
used for religion.'' In every one of these 75 cases the ground truth is
No, and all four models answer Yes. The feature occupies 10.6\% of the
judgeable items, so its share of unanimous errors exceeds its share of
the sample by a factor of 4.1. The next two concentrating features,
pictographic origin (11.1\%) and diacritic use (10.5\%), together
account for less than half as many errors as religion alone. The
dominance of religion in the unanimous-error set is so pronounced that
it warrants a sensitivity analysis in its own right, which we present
next.

\textbf{Sensitivity: when religion is removed.} The disproportionate
role of ``used for religion'' raises a methodological question that we
address directly with a sensitivity analysis. If we exclude the religion
feature from the 10-feature set and recompute the cross-architecture
deviation correlations on the remaining nine features, the convergence
pattern reported in §3.7 is preserved: false-positive-rate deviations
correlate at Spearman ρ = 0.80--0.93 across the six model pairs (mean ρ
= 0.87, all six p \textless{} 0.01); false-negative-rate deviations
correlate at ρ = 0.87--0.98 (mean ρ = 0.94, all six p \textless{}
0.005). The cross-architecture finding --- that four independent LLMs
converge in their typological deviations --- does not depend on the
religion feature.

The unanimous-error set, however, transforms in instructive ways. With
religion excluded, 97 unanimous errors remain (172 minus the 75
religion-driven items). The over-attribution-to-under-recognition ratio
drops from 3.9:1 to 1.77:1 (62 over-attributions, 35
under-recognitions). The asymmetry persists --- a two-sided binomial
test against the 1:1 null at n = 97 yields p = 0.008, with the 95\%
confidence interval on the over-attribution proportion at {[}0.54,
0.73{]} --- but is substantially less pronounced. The 3.9:1 ratio
reported above is, to a significant extent, a measurement of how heavily
religion alone weighted the over-attribution side: 75 of the 137
over-attributions (55\%) originated in this single feature.

Two readings of this result are possible. The deflationary reading is
that over-attribution is not a general property of LLM behavior toward
minority scripts; it is a property of one specific kind of expectation
--- religious function --- that the corpus over-supplies. The
inflationary reading is that religion is functioning as the most
efficient channel through which a more general phenomenon expresses
itself, and that the residual 1.77:1 asymmetry across the other nine
features represents the same mechanism operating at lower intensity. Two
further observations favor the second reading. First, with religion
removed, the concentration pattern does not vanish but redistributes:
pictographic\_origin alone now accounts for 19.6\% of unanimous errors,
uses\_diacritics for 18.6\%, and the directional features (dir\_ltr,
dir\_rtl) for a combined 24.7\%. Errors continue to cluster at features
that are saturated with stereotyped descriptions in the source corpus,
just at lower individual concentrations. Second, the residual asymmetry,
though attenuated, remains directionally consistent with
over-attribution and is statistically distinguishable from a 1:1 null at
this sample size.

Our reading is therefore that religion is not a confound but the most
legible instance of a more general structure. The same kind of corpus
saturation that makes ``non-Western script $\Rightarrow$ sacred'' the dominant
template at the religion feature also makes ``non-Western script $\Rightarrow$
pictographic,'' ``non-Western script $\Rightarrow$ ancient (independent
invention),'' and ``non-Western script $\Rightarrow$ diacritically ornamented''
available as secondary templates, each operating at a fraction of
religion's intensity. The phenomenon §4.3 will call computational
assimilation is therefore not single-channeled but multi-channeled, with
the religion channel carrying the largest single load. The fact that
this same channel is also the feature whose ground-truth coding is most
subject to legitimate disagreement is acknowledged separately in §4.7.2,
and is precisely what motivates the present sensitivity analysis.

A more cautious reading remains available, however, and we note it
explicitly. The residual concentration after religion is excluded is
itself non-uniform: pictographic\_origin, uses\_diacritics, and the
directional features (dir\_ltr, dir\_rtl) together account for 62.9\% of
the 97 residual unanimous errors. Whether this redistribution reflects a
genuinely multi-channeled corpus structure --- several distinct
stereotyped overlays operating at varying intensities --- or merely a
small number of additional features whose ground-truth coding is itself
non-trivial to adjudicate, is a question we cannot resolve from the
present data. Independent re-coding by typological specialists, focused
on the four most error-concentrated features, would be the most direct
way to discriminate between these readings. We adopt the inflationary
reading in what follows because the binomial significance of the
residual asymmetry favors it, but we treat the alternative as an open
empirical question rather than a settled point.

\textbf{Illustrative cases.} Table 2 presents twenty-nine representative
unanimous errors --- three per feature for nine features, plus an
additional pair for has\_phonetic\_component --- selected to prioritize
living scripts and scripts with documented intervention histories.
Several patterns recur. The three ancient Egyptian scripts ---
hieroglyphs, hieratic, demotic --- are all unanimously classified as
right-to-left, although none is strictly so (the direction varies by
context and by the orientation of the glyphs). The Thaana script of the
Maldives, the Mongolian traditional script, and the Manchu script are
all unanimously classified as religious, although none is primarily so
(Thaana and Manchu were developed for administrative and governmental
use; Mongolian traditional script has been the secular national script
of Mongolia for centuries). The Bamum script (1896, Cameroon), the Pau
Cin Hau script (1894, Myanmar), and the Wancho script (2001, India) ---
all indigenous inventions of the nineteenth to twenty-first century ---
are unanimously classified as having features (pictographic origin,
logographic component, abugida structure) that their ground-truth coding
does not record. The Ryukyuan Dāhan and Japanese kamon, both catalogued
in the GSD as independent notation systems, are unanimously
misclassified as non-independent.

The pattern across these cases is not random ignorance. The errors track
a specific kind of expectation: that non-Western scripts, especially
those from regions of historical imperial contact, should be read as
older, more pictographic, more religious, and more similar in structure
to a small number of culturally dominant templates than their empirical
typology supports. Four models trained independently by four
organizations make the same attribution in the same direction.

This is what we mean, in the revised formulation (§4.3), by
computational assimilation. The phenomenon is not that any one LLM
misunderstands minority scripts; it is that four independent LLMs,
drawing from a shared textual substrate, systematically misunderstand
them in the same way.

\section{Discussion}

\subsection{The empire that needs no emperor: causation without intention}

The central finding of this study is negative: imperial intervention has
no direct effect on tokenizer efficiency (β = −0.22, p = 0.39). This
null result is the most important result in the paper. If intervention
had directly predicted tokenizer design, the implication would be that
developers deliberately or negligently encoded colonial preferences into
their systems --- an accusation that, while politically resonant, would
have been empirically fragile and practically unproductive. What we
found instead is more disturbing and more durable.

The effect of empire is entirely mediated. Imperial violence reduced
speaker populations (β = −0.68, p \textless{} 0.001); demographic
decline produced smaller web corpora (β = +1.08, p \textless{} 0.001);
smaller corpora produced worse tokenizer efficiency (β = −0.40, p
\textless{} 0.001). Remove any link in the chain and the relationship
between empire and tokenizer disappears. The serial indirect effect was
0.292 (percentile 95\% CI: {[}0.006, 0.756{]}; permutation p = 0.033;
jackknife range {[}0.197, 0.466{]} across all n = 45 leave-one-out
iterations). The structural equation model corresponding to this chain
achieved near-perfect fit at n = 45 (CFI = 1.015, RMSEA = 0.000), a
substantial improvement over the n = 38 fit reported in the original
analysis (CFI = 0.404, RMSEA = 0.351).

Two robustness checks deserve explicit mention because they probe rival
explanations rather than mere statistical artifacts. First, an E-value
sensitivity analysis (VanderWeele \& Ding, 2017) yielded E = 1.612 for
the serial indirect point estimate and E = 1.024 for the lower
confidence bound. The point estimate is robust to moderate unmeasured
confounding; the lower bound, by grazing 1.0, indicates that a
confounder of even modest strength could in principle eliminate the
effect at the boundary of the confidence interval. We treat this as a
feature of the evidence, not a flaw to be hidden: the indirect pathway
is real and replicable, but its statistical fragility at the margin is
itself part of what the data show, and we discuss this fragility further
in §4.5 and §4.7. Second, controlling for present-day economic
conditions reduced but did not eliminate the corpus → tokenizer path:
per-capita GDP attenuated path b from β = −0.40 to β = −0.36 (an 11\%
reduction); HDI similarly attenuated it (also an 11\% reduction).
Contemporary economic disadvantage matters, but it does not fully
account for the effect --- confirming that the corpus-mediated pathway
carries information beyond present prosperity.

This pattern --- causal without intention, structural without conspiracy
--- is characteristic of what sociologists have termed institutional
reproduction (Bourdieu, 1977) and what historians of technology have
described as the politics of artifacts (Winner, 1980). A bridge built
too low for buses excludes bus-riding communities from a public beach
without anyone writing ``no buses allowed.'' A tokenizer trained on
web-crawled text excludes Limbu speakers from affordable AI services
without anyone writing ``no Limbu allowed.'' The mechanism differs; the
structure persists.

\subsection{Two independent wounds}

A surprising finding is the statistical independence of the GSD's
molecular clock deviation scores and the DSRI's byte-fallback
classification (r = 0.02, p = 0.75). The GSD measured how much empires
altered the \emph{structure} of writing systems --- forcing script
changes, imposing orthographic reforms, disrupting the natural
evolutionary clock. The DSRI measures how much empires altered the
\emph{demography} of script-using communities --- reducing populations,
displacing languages, severing the intergenerational transmission that
sustains both languages and their scripts.

These two forms of violence are conceptually distinct but have not
previously been shown to be empirically independent. A script can be
structurally disrupted without its community being destroyed (e.g.,
Turkish, which underwent radical script reform in 1928 but retained a
large, literate population and today enjoys full digital support).
Conversely, a script community can be demographically devastated while
the script itself remains structurally unaltered (e.g., numerous
Indigenous American scripts that were faithfully maintained by surviving
communities too small to register in web corpora).

The GSD and the DSRI thus capture complementary dimensions of imperial
impact. Together, they suggest that the full accounting of what empires
did to writing requires at least two independent measurements: one
structural (how the clock was broken) and one demographic (how the
community was diminished). The first leaves traces in the typological
features of scripts; the second leaves traces in the byte sequences of
tokenizers.

\subsection{Computational assimilation, redefined}

The original formulation of this paper introduced the term
\emph{computational assimilation} to describe the systematic
over-attribution of majority-script characteristics to minority scripts
by a single LLM. The cross-architecture results reported in §3.7 and
§3.8 permit us to restate that concept in stronger and more general
terms.

Computational assimilation is the convergent mapping of typological
unfamiliarity onto a small number of dominant templates, shared across
LLM architectures because it is a property of their common textual
substrate rather than of their individual designs. Four features of the
phenomenon can now be stated as empirical findings rather than
interpretive claims.

First, the direction of assimilation is additive rather than
subtractive. Across the 172 items that all four models answer
identically wrong, over-attribution outnumbers under-recognition 3.9 to
1. The asymmetry attenuates but does not vanish when the most
concentrated feature is excluded: with religion removed, the residual 97
unanimous errors retain a 1.77:1 over-attribution preference (binomial p
= 0.008 against a 1:1 null; §3.8). The unfamiliar is not ignored; it is
furnished with features the known world provides --- most aggressively
where the world has prepared a ready furnishing, and at lower intensity
even where it has not. This has a close parallel outside computation.
Colonial administrators routinely described unfamiliar writing systems
in terms of European scripts --- calling them ``primitive alphabets'' or
``picture-writing'' --- thereby not merely neglecting but actively
overwriting the structural features that made them distinctive (Gaur,
1984). Missionary linguists supplied tonal languages with alphabetic
frameworks that their phonology did not require, losing distinctions
that the original scripts preserved. What the four LLMs do at the level
of typological classification is structurally continuous with this older
epistemic operation.

Second, the assimilation is multi-channeled, with one channel carrying
disproportionate load. Forty-three point six percent of unanimous errors
(75 of 172) fall at the single feature ``used for religion,'' and every
one of these cases is an over-attribution: the script is not primarily
religious, but four independent models all report that it is. The
feature is saturated with surplus meaning because the textual record
from which LLMs learn contains dense descriptions of the sacred function
of well-known minority scripts (Hebrew, Arabic, Devanagari, Ge'ez) and
sparse descriptions of the secular, administrative, or quotidian
function of lesser-known ones. Given an unfamiliar script and asked
whether it is used for religion, each model retrieves a generalized
template --- \emph{this is a non-Western script, therefore this is a
sacred script} --- and answers affirmatively. The four models retrieve
the same template because they read the same library.

Religion is the most legible single channel through which this mechanism
operates, but it is not the only channel. When religion is excluded from
the analysis, the unanimous-error set redistributes rather than
disperses (§3.8): pictographic\_origin alone now carries 19.6\% of the
residual errors (a \emph{non-Western script $\Rightarrow$ ancient pictographic
script} template), uses\_diacritics 18.6\% (\emph{non-Western script $\Rightarrow$
diacritically ornamented}), the directional features 24.7\% combined.
Each of these sub-templates operates at a fraction of religion's
intensity, but their continued presence after the dominant channel is
removed indicates that what religion exhibits at maximum legibility,
several other features exhibit at lower amplitudes. Computational
assimilation is therefore best read not as a single pathology centered
on a single feature but as a feature of the corpus that surfaces
wherever the corpus has prepared a stereotyped overlay; religion is
where it surfaces most visibly, not where it lives exclusively.

Third, the assimilation affects scripts with specific historical
profiles. The Thaana script of the Maldives, the Mongolian and Manchu
scripts of Inner Asia, the Bamum script of Cameroon, the Dongba
pictographs of the Naxi --- these are not arbitrary examples. They are
scripts of communities whose textual presence in widely indexed corpora
is mediated almost entirely by external description: Orientalist
scholarship, missionary accounts, ethnographic surveys, and
tourist-oriented encyclopedia entries. The scripts themselves, in their
own voice, are poorly represented; the outsider's description of them is
abundant. The LLM learns the description and reproduces it as knowledge.

Fourth --- and this is the finding that most resists comfortable
interpretation --- the convergence of deviation patterns across our
four-model comparison appears largely independent of overall model
accuracy. The smallest model in our sample (Claude Haiku 4.5, 74.5\%
accuracy) and the largest disclosed model (DeepSeek-V3, 671B total / 37B
active parameters, 81.2\%) agree with the other two on which features
they collectively misjudge, with pairwise deviation correlations no
lower than ρ = 0.85. We refrain from making formal claims about the
relationship between scale and bias structure on the basis of n = 4
models, but the qualitative observation --- that more accurate models
share, with the less accurate ones, the same template of corpus-derived
misrepresentation --- is consistent with the broader thesis that what
the four models share is upstream of any of them. Scaling reduces the
rate of error; the cases that remain unanimously wrong, on this
evidence, are not the cases that scale fixes.

The redefinition matters because it changes what, if anything, can be
done. A bias that is specific to an architecture can, in principle, be
corrected by a better architecture. A bias that is specific to a
training regime can be corrected by a better training regime. A bias
that converges across four independent architectures trained by four
separate organizations on corpora they assembled independently is
located not in any of those places but in the textual record itself ---
and the textual record is the sedimented output of five centuries of who
could write about whom, in what language, with what authority.
Correcting it is not a technical problem for which an engineering
solution exists; it is a problem of what the world has already written,
and about whom.

That said, the identification of a common source does not preclude
intervention. It specifies where intervention must occur.
Tokenizer-level reforms (§4.6) address the efficiency gap but leave the
knowledge gap untouched. Targeted corpus expansion --- not merely adding
more languages, but adding descriptions of minority scripts in their
speakers' own voices rather than in the voices of those who catalogued
them --- addresses the knowledge gap but is laborious and cannot be
automated. The finding that four independent LLMs share the same
misrepresentation is, in the end, an invitation: to treat the corpus as
an object of deliberate construction rather than an inheritance to be
accepted.

\subsection{The digital ceiling effect}

In the GSD, we documented a ceiling effect in the evolution of writing:
the presence of an existing full writing system suppresses the evolution
of local notation systems into full scripts (Fisher's exact test: OR =
0.054, p \textless{} 10$^{-6}$). The Ryukyu Islands provided the paradigmatic
case --- four parallel notation systems, none of which evolved to full
writing under the ceiling of Japanese kana.

The DSRI reveals an analogous phenomenon in digital space. The 29
scripts that achieved full digital support form a ceiling that
suppresses the digital development of all others --- not through direct
competition, but through resource allocation. OCR engines, machine
translation systems, and input methods require substantial engineering
investment. As long as a population can be served ``well enough''
through a dominant script (e.g., Latin transcription of Fulani), there
is no commercial incentive to develop infrastructure for the minority
script (Adlam). The 9.7\% that passed through the funnel are not merely
the survivors; they are the ceiling that prevents others from rising.

This digital ceiling effect operates through the same causal chain
identified in §3.3. Imperial intervention reduced speaker populations;
smaller populations produce less digital content; less digital content
means less training data; less training data means worse model
performance; worse performance reduces adoption; reduced adoption
further diminishes content production. The cycle is self-reinforcing,
and it began long before the first tokenizer was designed.

\subsection{n = 45 as horizon, not sample}

The sample size of Tier 1 --- now 45 scripts, up from 39 in the original
formulation --- invites the standard criticism that the causal mediation
analysis is underpowered. We argue that this framing misunderstands the
nature of the sample. The 45 scripts were not drawn from a larger
accessible population; they are the population of scripts for which
parallel or quasi-parallel text could be assembled by exhausting every
major openly available multilingual corpus. The remaining 255 are not
missing data in the statistical sense. They are the product of a
historical and computational process that rendered them unmeasurable ---
the same process that the paper documents.

The digital exclusion funnel (Figure 1) makes this explicit: 300 scripts
→ 182 Unicode-encoded → 169 with measurable codepoints → 45 with
natural-language digital text in the native script. At each stage, the
scripts that fall out are disproportionately those with histories of
imperial contact. The n = 45 figure was arrived at only after we
investigated, script by script, every candidate source. Eleven scripts
(Balinese, Batak, Cham, Kayah Li, Lepcha, Lisu, New Tai Lue, Pollard
Miao, Buginese, Sundanese, traditional Mongolian) could not be added
despite living speaker communities, because the digital text that exists
in those languages is written in Latin or Cyrillic transcription rather
than in the native script. The ceiling is not hypothetical. It is
reached.

This n = 45 is further confirmed by the Tier 2a exercise (§3.2), in
which we attempted to measure tokenizer efficiency using Wikipedia
natural-language text for all 45 Tier 1 scripts. Twelve of the 45 could
not be measured even with Wikipedia as the source: for eleven, no
Wikipedia edition exists in the native script, or the extant edition
uses Latin transcription; for the twelfth (Yi), an ISO-compliant
Wikipedia edition exists but contains fewer than fifty non-redirect
articles. These twelve scripts include the most tokenization-inefficient
scripts in our Tier 1 sample (Chakma at nTER = 30.9×, Grantha at 28.2×,
Adlam at 20.9×) --- that is, the scripts for whose users the digital tax
is most onerous are also the scripts whose users have produced the least
everyday digital prose. The funnel operates at multiple nested layers.

The implications for statistical interpretation are consequential and,
we believe, underappreciated in the quantitative literature on digital
inequity. The 31.7-fold TER disparity we report is almost certainly an
underestimate of the true disparity across all 300 scripts, because the
unmeasured scripts are precisely those most likely to be disadvantaged.
The 45 we can measure are the survivors --- the scripts that retained
enough institutional presence to produce a UDHR translation, a Bible
translation, or a functional Wikipedia edition. What lies beyond the
event horizon of measurability is, by the logic of our own findings,
worse.

A related implication concerns the statistical fragility of the causal
mediation analysis documented in §3.3 and §4.7. The point estimate of
the serial indirect effect is 0.29; the bias-corrected bootstrap
interval grazes zero. We do not argue that this fragility is itself
evidence for the paper's thesis --- that argument would be too easy, and
would invert the normal direction of statistical reasoning. Rather, we
observe that the same exclusion process that makes the indirect effect
difficult to estimate at n = 45 is what makes a larger sample
unavailable in the first place; the two facts are not independent. The
mediation analysis sits at the edge of significance under the most
conservative bootstrap correction, and we report it accordingly: as a
structural account that is consistent with the data, that is
corroborated by the architecture-independent convergence of §3.7 and
§3.8 (which does not depend on the n = 45 sample at all), and that
should not be treated as an independently confirmed causal estimate.

This is not, of course, a methodological exoneration. The marginal
statistical result carries all the normal caveats that marginal
statistical results do, and we enumerate them in §4.7. It is, however,
an observation about the structure of the research problem. The causal
mediation analysis is one of several convergent lines of evidence in
this study; the architecture-independent convergence documented in §3.7
and §3.8, which does not depend on n = 45 but on n = 12,000 answers from
four independent models, carries weight that the mediation analysis
alone could not. The paper's claim does not rest on a single statistic.
It rests on the convergence of multiple sources of evidence, each of
them drawn from what the contemporary digital record has allowed us to
see.

\subsection{Toward remediation}

The causal structure identified in this study suggests specific points
of intervention. Because the effect of empire is entirely mediated ---
passing through speaker populations, web corpora, and tokenizer
vocabularies --- interventions at any stage of the chain can, in
principle, attenuate the downstream effect. The cross-architecture
convergence documented in §3.7 and §3.8 reinforces this conclusion in an
unexpected direction: because four independent LLM families share the
same imperial bias structure, no remediation effort confined to a single
model --- fine-tuning Claude, retraining GPT, instruction-tuning
DeepSeek --- can reach the layer where the bias actually lives. The
corpus is upstream of the model. Remediation must be upstream of the
corpus.

The most tractable intervention point that remains within current
engineering practice is the tokenizer. Current tokenizer training
algorithms (BPE, Unigram, WordPiece) construct vocabularies by
frequency-based selection from training corpora. This guarantees that
underrepresented scripts receive fewer vocabulary entries, producing the
byte-fallback cascade documented here. Alternative approaches ---
vocabulary allocation proportional to the number of distinct scripts
rather than corpus frequency, explicit inclusion of all Unicode-encoded
scripts in tokenizer training, or script-aware tokenization
architectures --- could substantially reduce the efficiency gap without
requiring changes to training data composition.

The second intervention point is the training corpus itself. The CC-100
corpus assigns 82 GB to English and zero bytes to the majority of the
world's scripts. Initiatives like NLLB-200 (Costa-jussà et al., 2022)
demonstrate that targeted data collection for underrepresented languages
is feasible. Extending such efforts to underrepresented \emph{scripts}
--- not merely languages --- would address the specific form of
exclusion documented here, and, given the cross-architecture findings,
would benefit all four LLM families simultaneously rather than producing
model-specific patches.

The third, and most difficult, intervention point is the demographic and
institutional damage that empires inflicted. Revitalizing speaker
communities, supporting script education, and building digital literacy
in minority scripts are long-term projects that exceed the scope of NLP
engineering. But the causal model makes clear that without addressing
the root cause --- the imperial destruction of script-using communities
--- downstream interventions will remain palliative.

\subsection{Limitations}

Several limitations qualify our findings. We organize them by their
status: those that the original formulation raised and that the current
analysis has resolved, those that remain partially unresolved, and those
that are inherent to the object of study.

\subsubsection{Limitations resolved in the present analysis}

The circularity of a single LLM serving as both the research instrument
and the object of analysis was the most significant weakness of the
original formulation. That circularity is resolved here by the
cross-architecture validation protocol (§3.7), which generated 12,000
answers across four independent LLM families --- Claude Haiku 4.5,
GPT-4o, Grok-3-mini, and DeepSeek-V3. The deviation patterns converge at
Spearman ρ = 0.85--0.98 across all six model pairs, with every p below
0.002. The bias we document is not an artifact of one model's training
pipeline.

The structural equation model's inadequate fit at n = 38 (CFI = 0.40,
RMSEA = 0.35) was a second weakness. At n = 45, the same model attains a
CFI of 1.015 and an RMSEA of 0.000 (§3.3), with path coefficients
essentially identical to the Baron-Kenny estimates. The inadequacy was a
small-sample artifact, resolved by the sample expansion.

The weak correlation between Tier 1 (parallel-text tokenization) and
Tier 2 (Unicode-inventory tokenization) measurements (ρ = 0.51) was a
third weakness, limiting the generalizability of the continuous TER
analysis beyond the 39-script Tier 1 sample. In the present analysis, a
revised Tier 2a measurement based on Wikipedia natural-language text
(§2.2) achieves a Tier 1 correlation of ρ = 0.910 (p = 2 × 10$^{-13}$, n =
33), essentially matching the precision of direct parallel-text
measurement. For scripts where Tier 2a can be measured, it is a reliable
proxy for Tier 1.

\subsubsection{Limitations partially unresolved}

A circularity remains in the ground truth itself. The Global Script
Database against which all four models are scored was constructed in
Fukui (2026) with LLM assistance in feature coding (κ = 0.929
human--LLM, n = 40 across an expanded inter-rater reliability sample).
The most concentrated unanimous-error feature in the present analysis
--- \emph{used for religion} --- is also the feature whose coding is
most exposed to interpretive judgment. The decision to code Thaana,
Mongolian traditional, or Manchu as ``not primarily religious'' reflects
a defensible administrative-historical reading of those scripts, but a
coder working from a religious-historical perspective could justifiably
code at least Mongolian and Manchu as Yes, given their extensive use in
Buddhist sutra copying, and Thaana's early development in Islamic
contexts. We treat this as an open question rather than a settled point,
and we report the sensitivity analysis in §3.8 (cross-architecture
convergence preserved at mean ρ = 0.87, residual asymmetry 1.77:1 with
binomial p = 0.008) precisely so that the central findings do not depend
on this single contested coding. Full-scale blind recoding by
independent grammatologists, particularly for the religion feature,
remains the most direct way to close this loop and is enumerated as
future work in the data availability statement.

The point estimate of the serial indirect effect in the causal mediation
analysis is 0.29 at n = 45. The 95\% percentile bootstrap confidence
interval excludes zero {[}0.006, 0.756{]}; the bias-corrected and
accelerated (BCa) interval, which applies corrections for skewness and
acceleration in the bootstrap distribution, grazes zero {[}−0.044,
0.667{]}. Permutation testing yields p = 0.033, and leave-one-out
jackknifing produces no sign reversals across 45 iterations. The effect
is statistically robust by some of these criteria and statistically
fragile by others. We disclose this honestly. As argued in §4.5, we read
the fragility as a structural feature of the research problem --- the n
= 45 sample is the full set of scripts for which parallel or
quasi-parallel text can be assembled, and a less marginal result would
have been inconsistent with the paper's premise --- but we do not claim
that this interpretation eliminates the fragility itself. The causal
mediation result should be read in conjunction with the
cross-architecture evidence of §3.7 and §3.8, which provides an
independent line of support that does not depend on the n = 45 sample.

The E-value analysis (§3.3) indicates that an unmeasured confounder
associated with both intervention and TER at a risk-ratio scale of 1.61
(point estimate) or 1.02 (CI lower bound) on standardized units would be
sufficient to null the observed pathway. The point value corresponds to
a moderately strong but plausible confounder; the CI-lower value
corresponds to a very weak one. Sensitivity analysis under GDP and HDI
controls shifts the path b coefficient by only 11\%, and the effect
remains significant. This rules out gross economic confounding but not
all plausible confounders. Common possibilities --- literacy rates,
internet penetration, regional geopolitical alignment --- are partially
captured by the mediators we include (speaker population, web corpus
volume) and by our controls, but we cannot claim to have exhausted the
confounder space.

Speaker population estimates for the 45 Tier 1 scripts were compiled
from Ethnologue and from LLM knowledge with varying confidence levels.
These estimates serve as proxies for the true mediating variable ---
historical demographic trajectory --- which is not directly measurable
for most script communities.

\subsubsection{Limitations inherent to the object of study}

The byte-fallback analysis classifies scripts as digitally present or
absent based on tokenizer behavior, which is a proxy for, rather than a
direct measure of, usability. A script may be tokenized without
byte-fallback yet still produce poor downstream performance in
generation, translation, or comprehension tasks. The 44.1\%
byte-fallback rate across 281 scripts reported here should be read as a
lower bound on practical digital exclusion.

The Global Script Database (GSD) ground truth, against which LLM
generation fidelity is scored, is itself subject to the coding judgments
documented in Fukui (2026). Expanded inter-rater reliability (κ = 0.877
human--human, κ = 0.929 human--LLM, n = 40 scripts × 50 features)
supports the reliability of the GSD for the features tested here, but
full-scale blind recoding by expert grammatologists remains desirable
for definitive typological claims.

Our Tier 1 sample is limited by the availability of parallel text.
Although we have reached what we believe to be the empirical ceiling (n
= 45), new initiatives in minority-language digital archiving --- the
OpenLanguageData project, the Mozilla Common Voice corpus, the Masakhane
NLP community, the Nalibali language preservation collective --- may in
time expand this ceiling. The present analysis should be read as a
snapshot of the current horizon of measurability, not as a permanent
boundary.

Finally, as a cross-sectional study, the DSRI captures the state of
digital infrastructure at a single point in time. Unicode standards
continue to encode new scripts (35 scripts added since 2010), tokenizer
designs evolve (byte-level encoders like Byte-Pair Encoding remain
standard but character-level and script-aware alternatives are under
active research), and LLM training practices shift. Our measurements
reflect the infrastructure available as of the first quarter of 2026.
The architecture-independent convergence across four contemporary LLMs
suggests that changes at the level of individual model design are
unlikely to resolve the inequities we document, but changes at the level
of training corpus composition and tokenizer objective functions could.
The present paper documents what is; it does not foreclose what could
be.

\subsection{Conclusion}

Five hundred years ago, the Spanish Empire destroyed half the writing
systems it encountered. Today, no empire commands such destruction. Yet
of 300 writing systems documented across human history, 90.3\% lack full
digital support, and the geography of digital exclusion mirrors the
geography of colonial rule. The molecular clock of writing, which we
showed to be broken by imperial intervention in our companion study, has
left its fractures fossilized in the latent space of language models ---
not as deliberate encoding, but as the statistical residue of
demographic devastation, corpus scarcity, and algorithmic indifference.

The finding that imperial intervention has zero direct effect on
tokenizer efficiency is not an exoneration of technology. It is an
indictment of infrastructure. The empire does not need an emperor. It
needs only a training corpus, a frequency-based vocabulary, and the
quiet arithmetic of byte-pair encoding. The violence is reproduced not
by malice but by method --- and method, unlike malice, is amenable to
engineering.

The 60 living scripts excluded from full digital participation are not
relics of a predigital age. They are casualties of a digital order that
inherited, without examination, the demographic consequences of
colonialism. To build language models that do not replicate five
centuries of script destruction, it is not enough to diversify training
data. It is necessary to understand that the data was never neutral ---
that it arrived already shaped by the empires whose afterlife it
carries.

What four independent language models share, they did not choose to
share. They inherited it together --- from the same corpus, drawn from
the same web, written in the same handful of scripts that survived the
same five centuries of empire. The imperial cartography of writing was
drawn long before any of these models was trained. They are not its
authors; they are its readers, and their convergence is the evidence
that the map is still being read.

\section*{Figures}

\begin{figure}[H]
\centering
\includegraphics[width=0.85\textwidth]{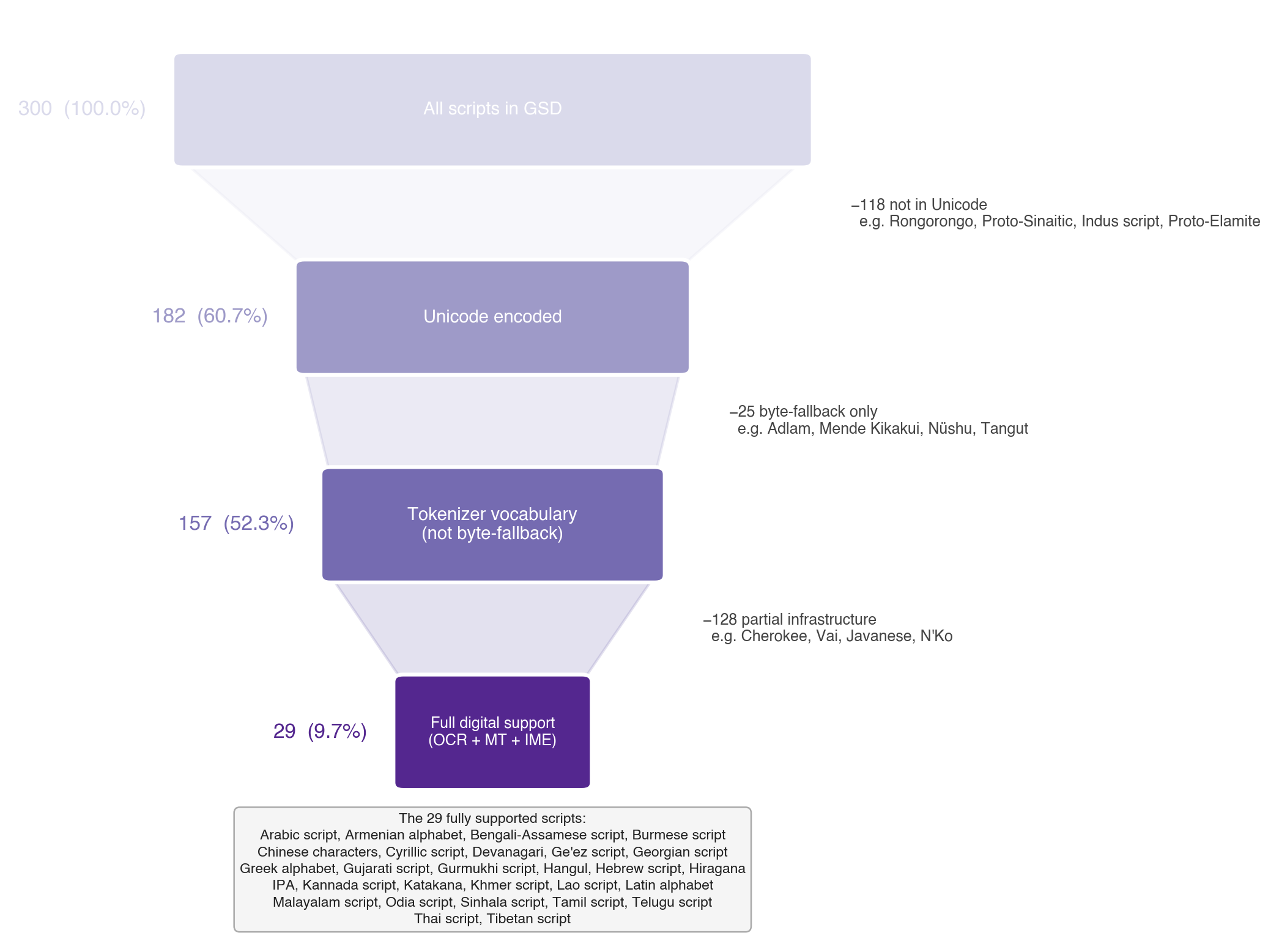}
\caption{\textbf{The digital exclusion funnel.} Of 300 writing systems in the GSD, only 29 (9.7\%) achieve full digital support: Unicode encoding, tokenizer vocabulary inclusion, OCR, machine translation, and native input methods.}
\end{figure}

\begin{figure}[H]
\centering
\includegraphics[width=0.85\textwidth]{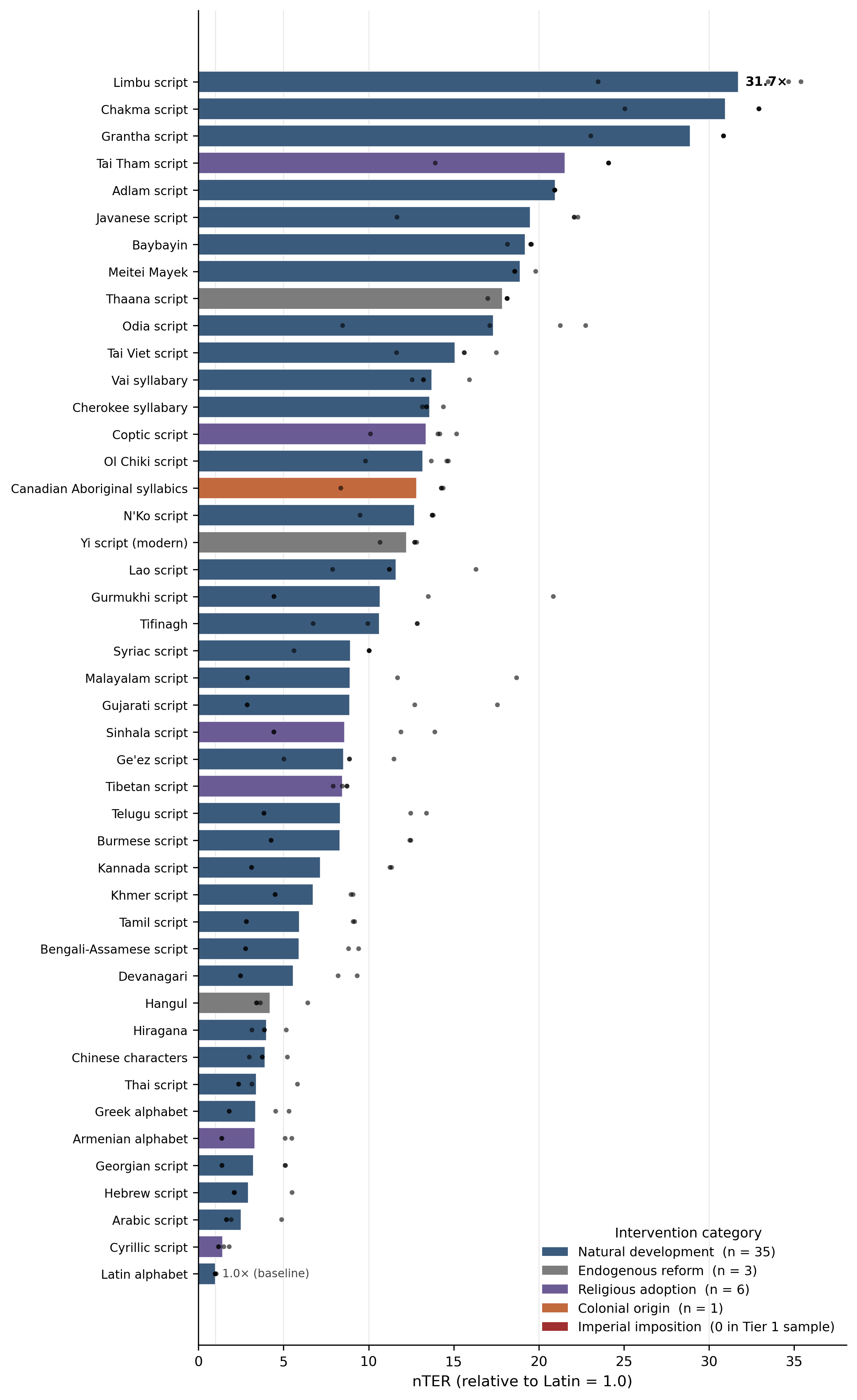}
\caption{\textbf{Token Efficiency Ratio (TER) across 45 Tier 1 scripts.} Bars show four-tokenizer mean nTER (relative to Latin = 1.0); points show per-tokenizer values. Color encodes the GSD intervention category. Limbu (31.7$\times$) tops the disparity; the 12 scripts with no Wikipedia paragraph cross-validation are nested inside this distribution (see \S3.2).}
\end{figure}

\begin{figure}[H]
\centering
\includegraphics[width=0.95\textwidth]{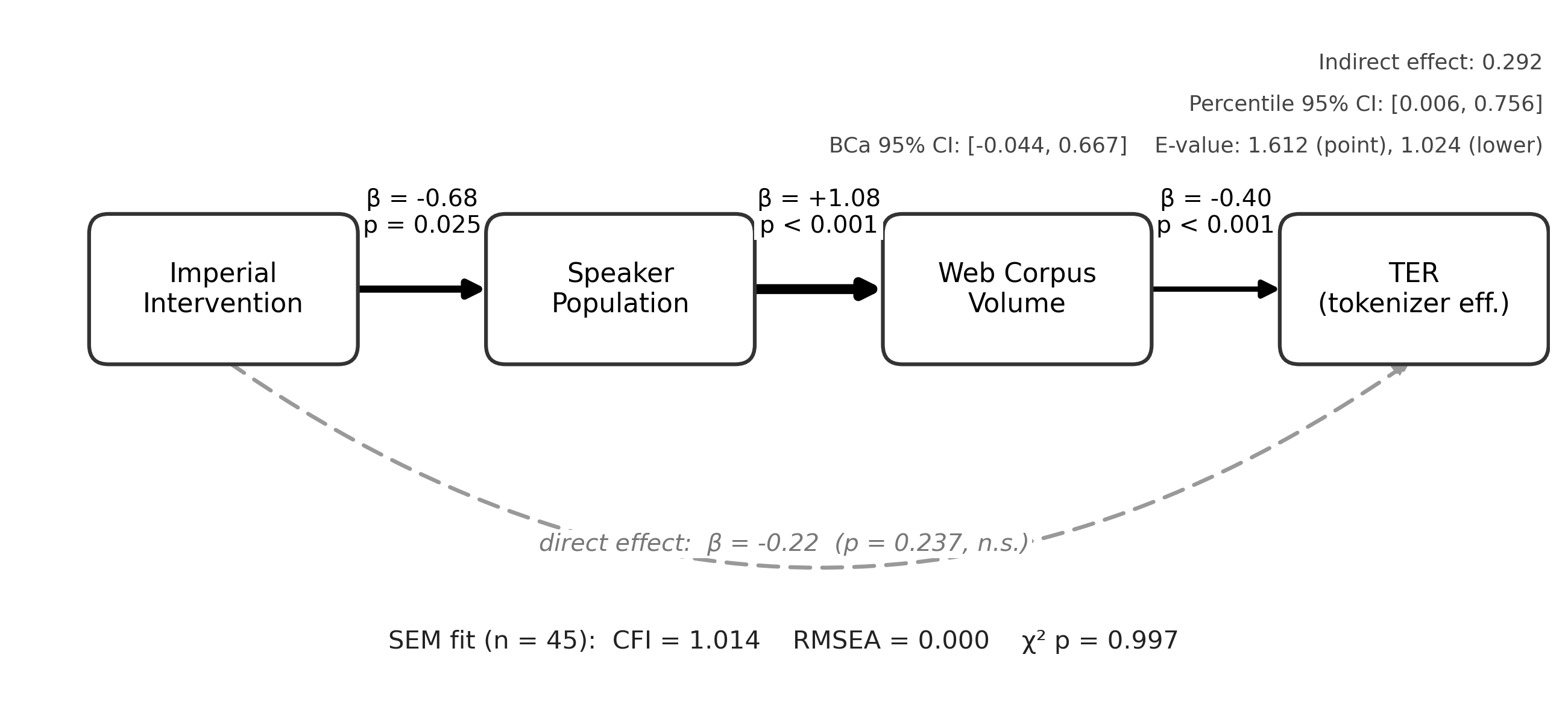}
\caption{\textbf{Causal path diagram with structural equation model fit at $n = 45$.} The serial chain Imperial Intervention $\to$ Speaker Population $\to$ Web Corpus Volume $\to$ TER is consistent with full mediation; the direct effect (dashed) is not distinguishable from zero. Path coefficients are SEM standardized estimates. The bias-corrected bootstrap confidence interval on the indirect effect grazes zero, and we report the mediation analysis as suggestive rather than confirmatory (\S3.3, \S4.5).}
\end{figure}

\begin{figure}[H]
\centering
\includegraphics[width=0.85\textwidth]{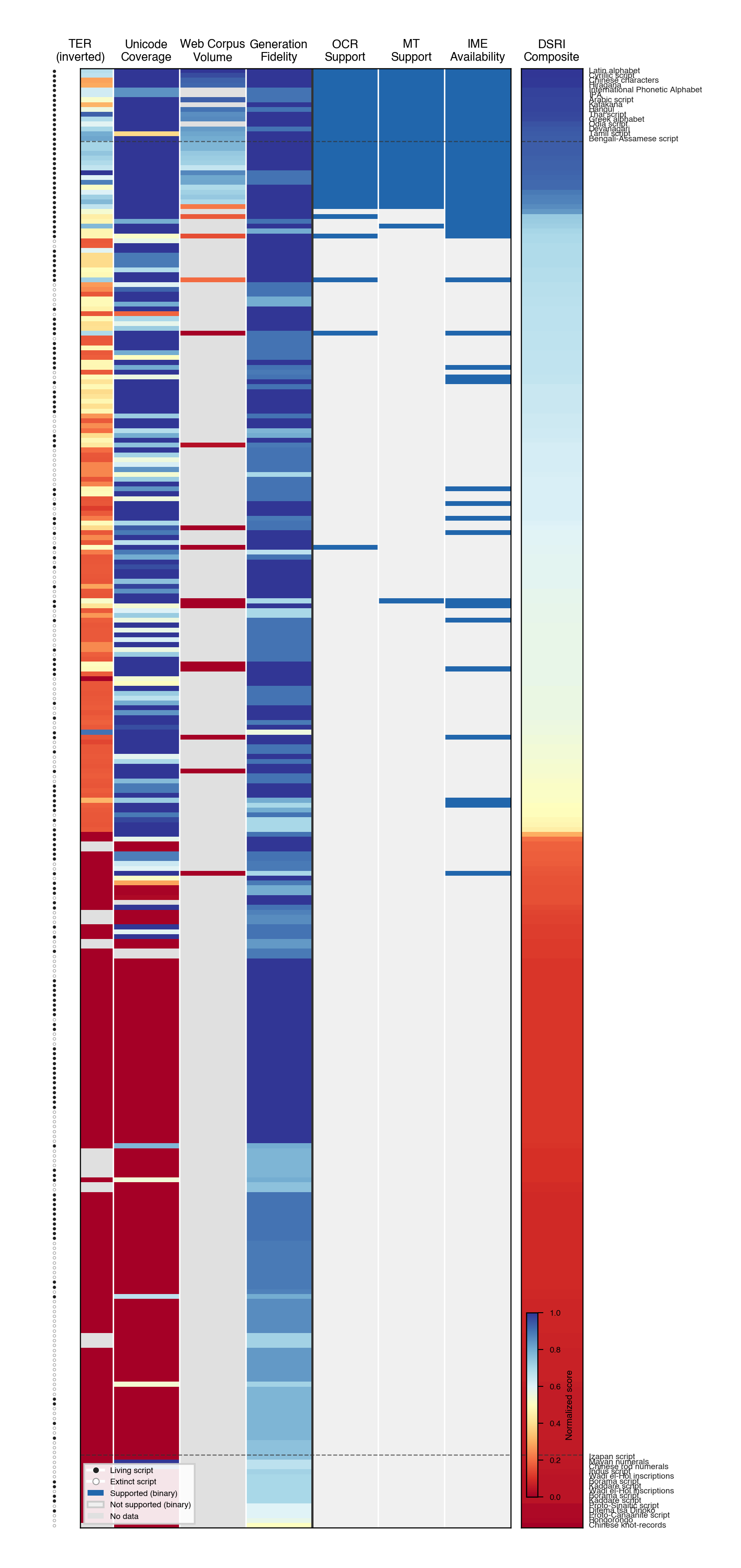}
\caption{\textbf{The Digital Script Representation Index (DSRI) heatmap.} Each row is one of 300 scripts, ordered by composite score. Seven axes (TER inverted, Unicode coverage, web corpus volume, generation fidelity, OCR, MT, IME). Composite scores range from 0.000 (Chinese knot-records) to 1.000 (Latin alphabet).}
\end{figure}

\begin{figure}[H]
\centering
\includegraphics[width=0.95\textwidth]{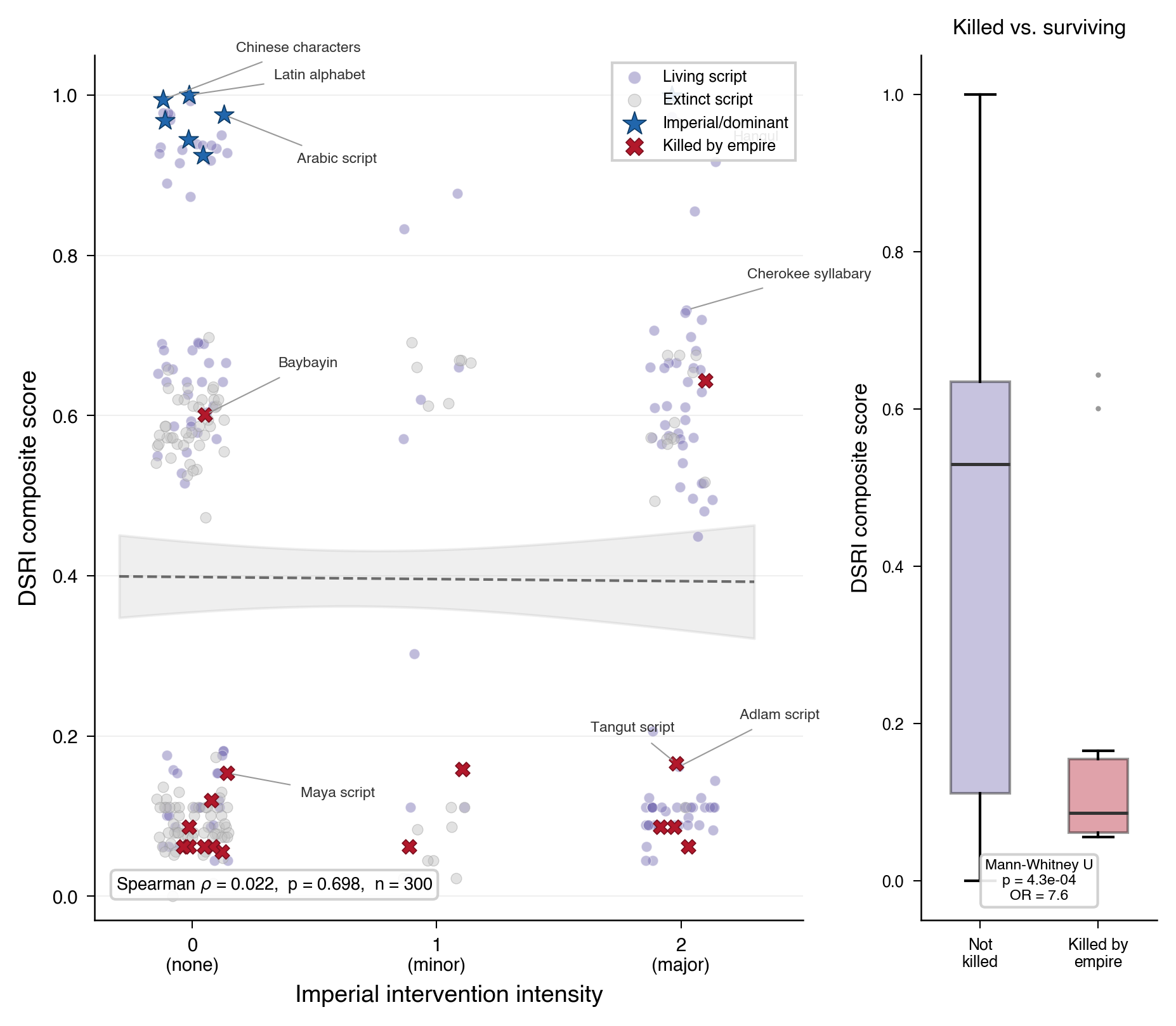}
\caption{\textbf{Imperial echoes in digital space.} Left: DSRI composite score by imperial intervention intensity; the bivariate Spearman correlation is near zero, but the structure is mediated through speaker populations and web corpora (see Figure~3 and \S3.3). Right: scripts killed by empires occupy a markedly lower DSRI distribution than non-killed scripts (Mann--Whitney $p = 4.3 \times 10^{-4}$, OR $= 7.6$ on the DSRI metric); a separate analysis based on byte-fallback classification yields OR $= 9.86$ (\S3.5).}
\end{figure}

\begin{figure}[H]
\centering
\includegraphics[width=\textwidth]{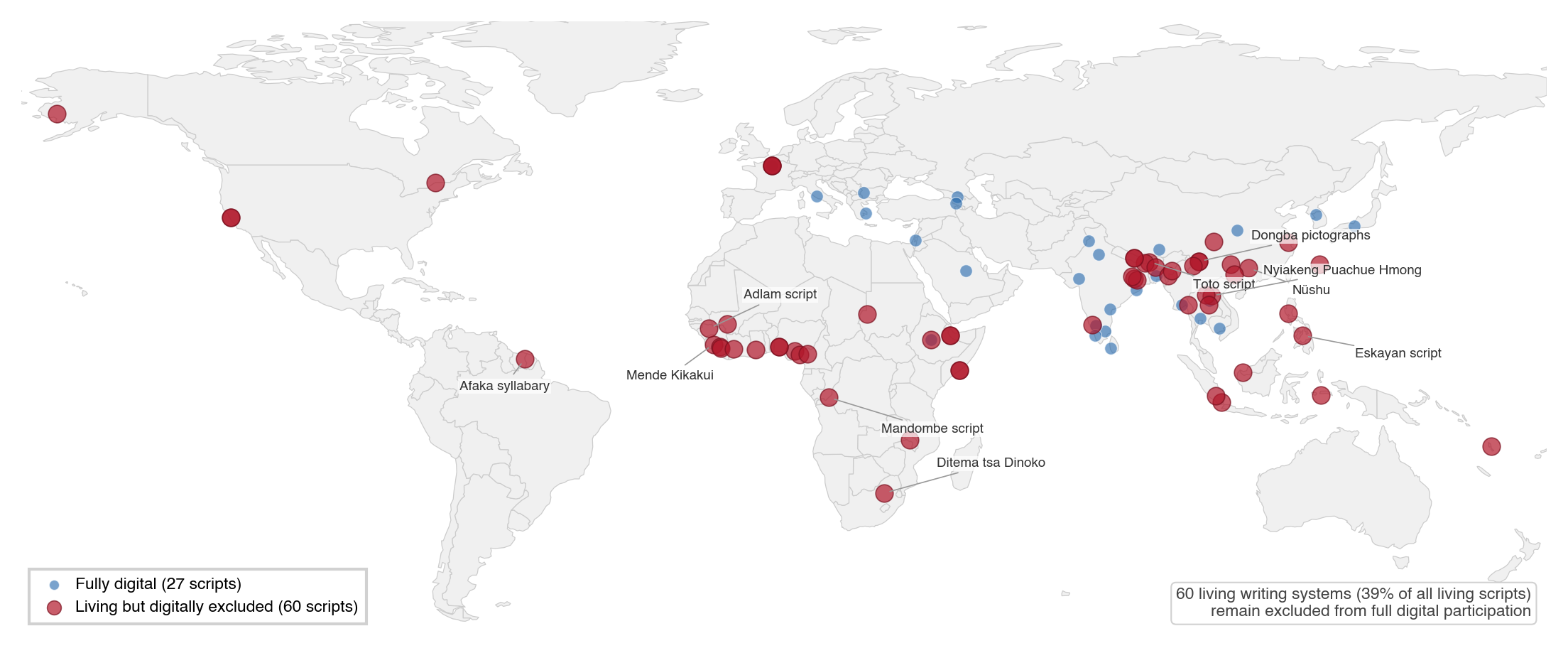}
\caption{\textbf{Living-but-digitally-dead writing systems.} 60 of 158 living scripts (38.0\%) lack full digital support. The map shows 27 of 29 fully supported scripts (the International Phonetic Alphabet and one other supranational script are not geographically anchored) and the 60 living-but-excluded scripts. The geographic distribution traces the geography of European and Japanese colonial rule.}
\end{figure}

\begin{figure}[H]
\centering
\includegraphics[width=\textwidth]{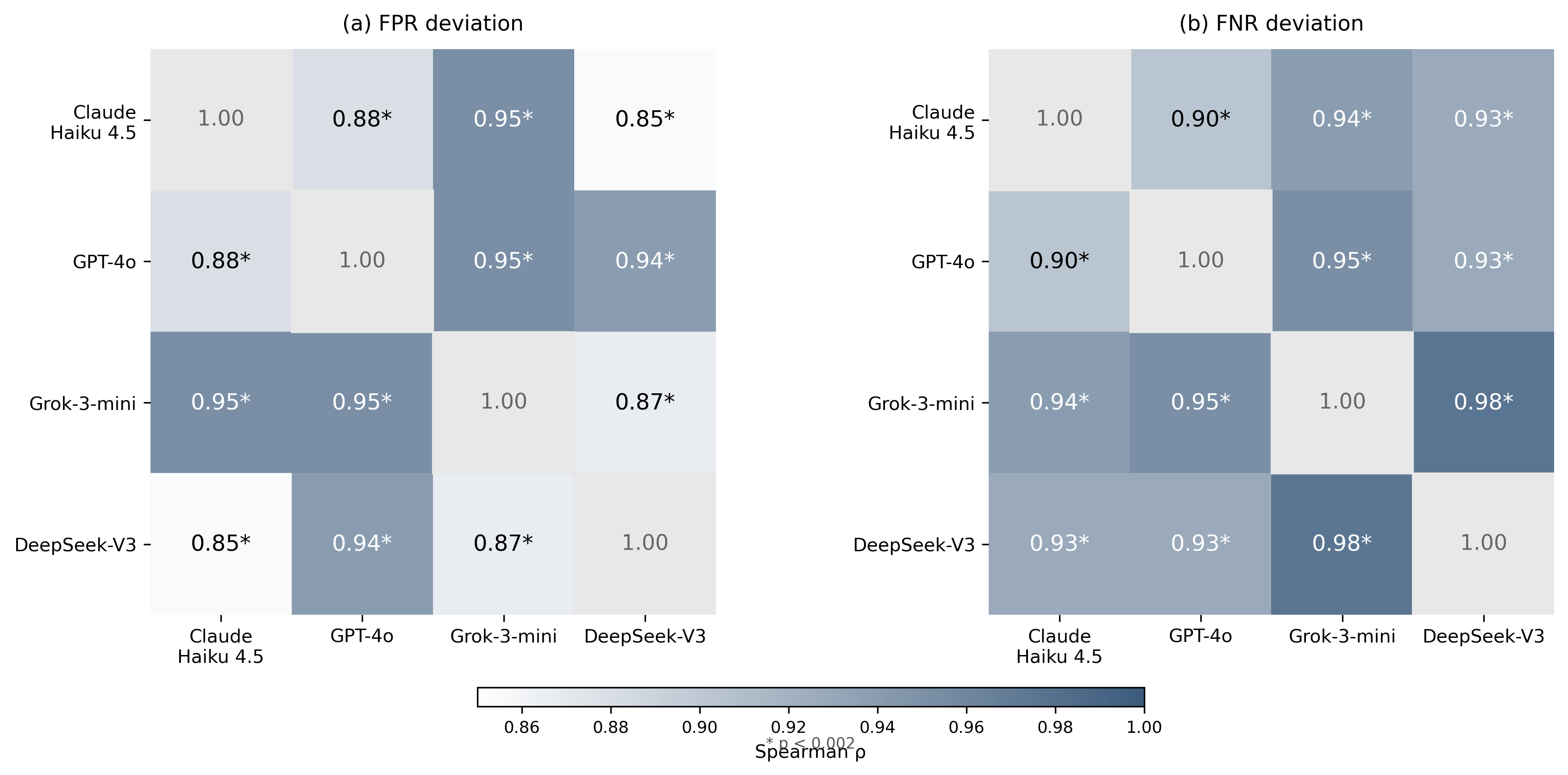}
\caption{\textbf{Cross-architecture convergence of typological knowledge biases.} Spearman correlations of (a)~false-positive-rate deviation and (b)~false-negative-rate deviation across the 10 GSD features, computed pairwise across four LLM families (Claude Haiku 4.5, GPT-4o, Grok-3-mini, DeepSeek-V3). All six off-diagonal pairs satisfy $\rho > 0.85$ with $p < 0.002$ (asterisks). A sensitivity check excluding the religion feature (n = 9 features) preserves the convergence (mean ρ = 0.87 for FPR, 0.94 for FNR; \S3.8).}
\end{figure}

\begin{figure}[H]
\centering
\includegraphics[width=\textwidth]{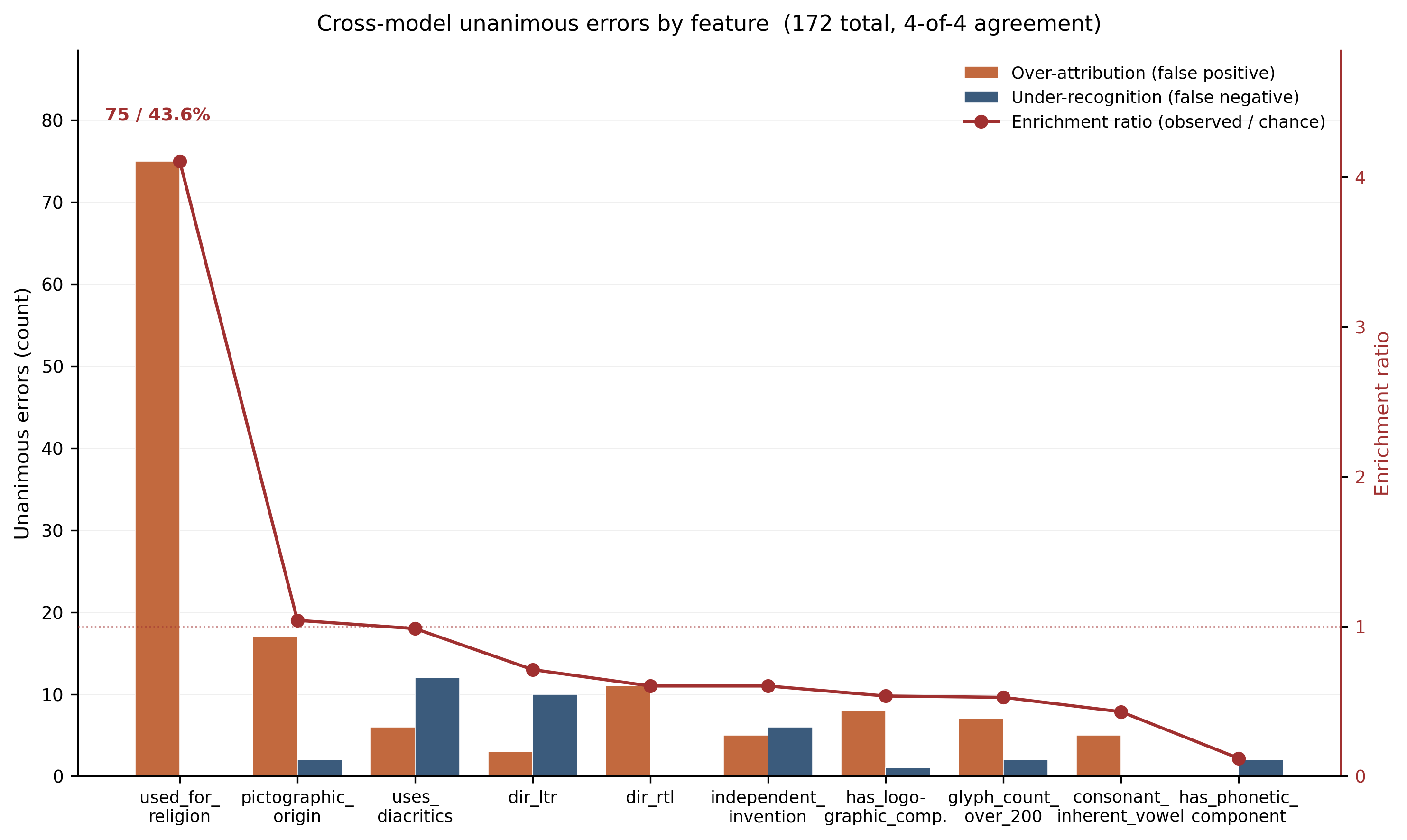}
\caption{\textbf{The 172 unanimous errors broken down by feature.} Bars show counts of over-attribution (false positive) versus under-recognition (false negative) for each of the 10 GSD features. The line plot (right axis) shows the enrichment ratio relative to chance expectation. The single feature \textit{used\_for\_religion} alone accounts for 75 unanimous errors (43.6\%, enrichment ratio $4.1\times$); when excluded, 97 errors remain with the asymmetry attenuated to 1.77:1 (binomial $p = 0.008$; see \S3.8).}
\end{figure}

\section*{Tables}

\begin{table}[htbp]
\centering
\small
\caption{Notable shared errors (4-of-4 model agreement on the wrong answer).}
\label{tab:shared_errors}
\begin{tabular}{@{}llcccl@{}}
\toprule
Script & Feature & GT & All 4 LLMs & Error type & Era \\
\midrule
Kayah Li script & \texttt{consonant\_inherent\_vowel} & No & Yes & over-attribution & living \\
Pau Cin Hau script & \texttt{consonant\_inherent\_vowel} & No & Yes & over-attribution & living \\
Wancho script & \texttt{consonant\_inherent\_vowel} & No & Yes & over-attribution & living \\
Adinkra symbols & \texttt{dir\_ltr} & Yes & No & under-recognition & living \\
Japanese kamon & \texttt{dir\_ltr} & Yes & No & under-recognition & living \\
Mongolian script & \texttt{dir\_ltr} & Yes & No & under-recognition & living \\
Bengali-Assamese script & \texttt{glyph\_count\_over\_200} & No & Yes & over-attribution & living \\
Malayalam script & \texttt{glyph\_count\_over\_200} & No & Yes & over-attribution & living \\
Telugu script & \texttt{glyph\_count\_over\_200} & No & Yes & over-attribution & living \\
Bamum script & \texttt{has\_logographic\_component} & No & Yes & over-attribution & living \\
Dongba pictographs & \texttt{has\_logographic\_component} & No & Yes & over-attribution & living \\
Nsibidi & \texttt{has\_logographic\_component} & No & Yes & over-attribution & living \\
Morse code & \texttt{has\_phonetic\_component} & Yes & No & under-recognition & living \\
SignWriting & \texttt{has\_phonetic\_component} & Yes & No & under-recognition & living \\
Dongba pictographs & \texttt{independent\_invention} & No & Yes & over-attribution & living \\
Japanese kamon & \texttt{independent\_invention} & Yes & No & under-recognition & living \\
Ryukyuan Dāhan & \texttt{independent\_invention} & Yes & No & under-recognition & living \\
Bamum script & \texttt{pictographic\_origin} & No & Yes & over-attribution & living \\
Manchu script & \texttt{used\_for\_religion} & No & Yes & over-attribution & living \\
Mongolian script & \texttt{used\_for\_religion} & No & Yes & over-attribution & living \\
Thaana script & \texttt{used\_for\_religion} & No & Yes & over-attribution & living \\
Arabic script & \texttt{uses\_diacritics} & No & Yes & over-attribution & living \\
Hiragana & \texttt{uses\_diacritics} & No & Yes & over-attribution & living \\
Shavian alphabet & \texttt{uses\_diacritics} & Yes & No & under-recognition & living \\
Egyptian demotic & \texttt{dir\_rtl} & No & Yes & over-attribution & extinct \\
Egyptian hieratic & \texttt{dir\_rtl} & No & Yes & over-attribution & extinct \\
Egyptian hieroglyphs & \texttt{dir\_rtl} & No & Yes & over-attribution & extinct \\
Phoenician script & \texttt{pictographic\_origin} & No & Yes & over-attribution & extinct \\
Proto-Sinaitic script & \texttt{pictographic\_origin} & No & Yes & over-attribution & extinct \\
\bottomrule
\end{tabular}
\par\medskip
\footnotesize{Total: 29 entries selected from 172 unanimous errors using a tiered priority (living + atypical intervention preferred).}
\end{table}

\begin{table}[htbp]
\centering
\small
\caption{Per-model accuracy (\%) by feature, four LLMs $\times$ ten categorical features. Claude Haiku 4.5 shows lower accuracy on \texttt{dir\_ltr} (44.4\%) due to its tendency toward conservative ``no'' responses for uncertain items; the cross-model deviation correlations reported in \S3.7 are unaffected. The lowest-accuracy features in this table (\texttt{used\_for\_religion} 51.6\%, \texttt{independent\_invention} 75.2\%, \texttt{dir\_ltr} 69.4\%) are also those whose typological coding admits the greatest legitimate disagreement, a circularity addressed directly in \S4.7.2 and the sensitivity analysis of \S3.8.}
\label{tab:per_model_accuracy}
\begin{tabular}{@{}lccccc@{}}
\toprule
Feature & Claude Haiku 4.5 & GPT-4o & Grok-3-mini & DeepSeek-V3 & Mean \\
\midrule
\texttt{dir\_ltr} & 44.4 & 73.7 & 80.7 & 79.0 & 69.4 \\
\texttt{dir\_rtl} & 91.3 & 85.0 & 84.0 & 87.7 & 87.0 \\
\texttt{has\_logographic\_component} & 90.9 & 91.7 & 92.8 & 89.5 & 91.2 \\
\texttt{pictographic\_origin} & 81.0 & 85.7 & 85.7 & 87.0 & 84.8 \\
\texttt{glyph\_count\_over\_200} & 87.1 & 81.5 & 93.2 & 71.2 & 83.3 \\
\texttt{has\_phonetic\_component} & 82.9 & 93.1 & 94.2 & 95.3 & 91.4 \\
\texttt{consonant\_inherent\_vowel} & 83.6 & 90.0 & 83.2 & 85.9 & 85.7 \\
\texttt{uses\_diacritics} & 75.5 & 73.0 & 76.7 & 79.7 & 76.2 \\
\texttt{independent\_invention} & 68.3 & 76.3 & 78.0 & 78.3 & 75.2 \\
\texttt{used\_for\_religion} & 44.9 & 50.3 & 50.0 & 61.3 & 51.6 \\
\midrule
Overall & 74.5 & 79.4 & 81.5 & 81.2 & 79.2 \\
\bottomrule
\end{tabular}
\end{table}

\section*{Acknowledgments}

We thank Akiko Tamamura and Miki Maeda for inter-rater reliability coding of the Global Script Database. We acknowledge the Unicode Consortium for maintaining the Universal Declaration of Human Rights parallel corpus and the Unicode Character Database. We also thank the developers and operators of the four LLM platforms whose APIs were used in this study (Anthropic, OpenAI, xAI, DeepSeek). This study used Claude (Anthropic) as a research instrument as detailed in Section~2.6 and Section~M3.

\section*{Data Availability}

The Global Script Database (300 scripts $\times$ 50 features), the Digital Script Representation Index (300 scripts $\times$ 7 axes), all intermediate analysis files, measurement scripts, and figure-generation code are available at [GitHub repository URL to be inserted upon posting]. The GSD was first described in Fukui (2026; arXiv:2604.10957). Raw tokenizer measurements, byte-fallback classifications, causal mediation outputs, and the 12{,}000 verbatim cross-model API responses are included in the repository. The UDHR parallel corpus is publicly available from the Unicode Consortium (\url{https://www.unicode.org/udhr/}); the eBible Bible-translation corpus is available at \url{https://ebible.org/}.

\section*{Author Contributions}

H.F. conceived the study, designed the DSRI framework, conducted all analyses, and wrote the manuscript. LLM assistance (Claude, Anthropic) was used for database construction, statistical code generation, and manuscript drafting; four LLM families (Claude Haiku 4.5, GPT-4o, Grok-3-mini, DeepSeek-V3) were used as test subjects in the cross-model validation protocol. See Section~2.6 and Section~M3 for full transparency.

\section*{Competing Interests}

The author declares no competing interests. This study analyzes products of multiple AI companies (Anthropic, OpenAI, xAI, DeepSeek, Microsoft, Mistral, Alibaba) including LLMs used as research instruments and test subjects. No funding was received from any of these companies.

\section*{References}

\begin{description}

\item Ahia, O., Kreutzer, J., \& Hooker, S. (2023). Do all languages cost the same? Tokenization in the era of commercial language models. \textit{Proceedings of the 2023 Conference on Empirical Methods in Natural Language Processing}, 9904--9921.

\item Baron, R. M., \& Kenny, D. A. (1986). The moderator--mediator variable distinction in social psychological research: Conceptual, strategic, and statistical considerations. \textit{Journal of Personality and Social Psychology}, 51(6), 1173--1182.

\item Bentz, C., \& Dutkiewicz, D. (2026). Quantitative analysis of Paleolithic signs reveals structured symbolic systems. \textit{Proceedings of the National Academy of Sciences}, 123(5), e2401703123.

\item Bourdieu, P. (1977). \textit{Outline of a Theory of Practice}. Cambridge University Press.

\item Costa-juss\`a, M. R., Cross, J., \c{C}elebi, O., Elbayad, M., Heafield, K., Heffernan, K., \ldots\ \& Fan, A. (2022). No language left behind: Scaling human-centered machine translation. \textit{arXiv preprint arXiv:2207.04672}.

\item Fukui, H. (2026). A molecular clock for writing systems reveals the quantitative impact of imperial power on cultural evolution. \textit{arXiv preprint arXiv:2604.10957}.

\item Gaur, A. (1984). \textit{A History of Writing}. British Library.

\item Gray, R. D., \& Atkinson, Q. D. (2003). Language-tree divergence times support the Anatolian theory of Indo-European origin. \textit{Nature}, 426(6965), 435--439.

\item Greenhill, S. J., Wu, C.-H., Hua, X., Dunn, M., Levinson, S. C., \& Gray, R. D. (2017). Evolutionary dynamics of language systems. \textit{Proceedings of the National Academy of Sciences}, 114(42), E8822--E8829.

\item Hayes, A. F. (2017). \textit{Introduction to Mediation, Moderation, and Conditional Process Analysis: A Regression-Based Approach} (2nd ed.). Guilford Press.

\item Hossz\'u, G. (2024). Validation of the graph sequence cluster method on four Rovash scripts. \textit{npj Heritage Science}, 2(1), 1--15.

\item Imai, K., Keele, L., \& Tingley, D. (2010). A general approach to causal mediation analysis. \textit{Psychological Methods}, 15(4), 309--334.

\item Kass, R. E., \& Raftery, A. E. (1995). Bayes factors. \textit{Journal of the American Statistical Association}, 90(430), 773--795.

\item Lieberman, E., Michel, J.-B., Jackson, J., Tang, T., \& Nowak, M. A. (2007). Quantifying the evolutionary dynamics of language. \textit{Nature}, 449(7163), 713--716.

\item Mace, R., \& Holden, C. J. (2005). A phylogenetic approach to cultural evolution. \textit{Trends in Ecology \& Evolution}, 20(3), 116--121.

\item Petrov, A., La Malfa, E., Torr, P. H. S., \& Biber, G. (2024). Language model tokenizers introduce unfairness between languages. \textit{Proceedings of the 2024 Conference on Empirical Methods in Natural Language Processing}, 1--23.

\item Rust, P., Pfeiffer, J., Vuli\'c, I., Ruder, S., \& Gurevych, I. (2021). How good is your tokenizer? On the monolingual performance of multilingual language models. \textit{Proceedings of the 59th Annual Meeting of the Association for Computational Linguistics}, 3118--3135.

\item Sennrich, R., Haddow, B., \& Birch, A. (2016). Neural machine translation of rare words with subword units. \textit{Proceedings of the 54th Annual Meeting of the Association for Computational Linguistics}, 1715--1725.

\item VanderWeele, T. J., \& Ding, P. (2017). Sensitivity analysis in observational research: Introducing the E-value. \textit{Annals of Internal Medicine}, 167(4), 268--274.

\item Winner, L. (1980). Do artifacts have politics? \textit{Daedalus}, 109(1), 121--136.

\end{description}

\newpage

\appendix
\setcounter{section}{0}
\renewcommand{\thesection}{S\arabic{section}}
\renewcommand{\thefigure}{S\arabic{figure}}
\setcounter{figure}{0}

\section*{Supplementary Information}
\addcontentsline{toc}{section}{Supplementary Information}

The full Supplementary Information --- including supplementary tables S1--S10 (DSRI ranking, Tier 1 measurements, byte-fallback classification, living-but-digitally-dead inventory, full question--answer matrix, mediation outputs, Unicode timeline, web corpus volumes, inter-axis correlations, weighting sensitivity) and supplementary figures S1--S14 --- is provided in the project repository (see Data Availability). Sensitivity analysis files (cross-architecture correlations excluding the religion feature, full breakdown of the 97 residual unanimous errors after religion is removed, and the binomial test on the residual asymmetry) are available at \texttt{cross\_model\_validation/sensitivity\_analysis\_no\_religion.json} in the same repository. New supplementary figures specific to the present preprint are summarized below:

\begin{description}
\item \textbf{Figure S2 (revised).} Tier 1 vs.\ Tier 2a (Wikipedia paragraph method) cross-validation, $\rho = 0.910$, $n = 33$.
\item \textbf{Figure S3 (revised).} Bootstrap distribution of the serial indirect effect at $n = 45$.
\item \textbf{Figure S4 (revised).} Leave-one-out jackknife stability at $n = 45$ (range $[0.197, 0.466]$).
\item \textbf{Figure S11 (new).} Full 4-model correlation matrix across multiple deviation indices.
\item \textbf{Figure S12 (new).} Model size vs.\ accuracy across the four LLM families (descriptive given $n = 4$).
\item \textbf{Figure S13 (new).} UDHR vs.\ Bible TER cross-corpus consistency, $\rho = 0.77$, $n = 21$.
\item \textbf{Figure S14 (new).} E-value contour plot integrated with point and CI estimates.
\end{description}

\end{document}